\def\gapprox{{_>\atop{^\sim}}}
\def\lapprox{{_<\atop{^\sim}}}

\def\cmmt{\rm {cm^{-2}}}

\def\s-1{\rm {s^{-1}}}
\def\twco{$^{12}$CO}

\def\sskip{\vskip \baselineskip \noindent}

\def \0{\hbox to 0.5 em{}}
\def \etal{{\it et al.\ }}

\def \kms{km s$^{-1}\ $}

\def\myoversim#1#2{\lower2.0pt\vbox{\baselineskip0pt \lineskip0.2pt
    \ialign{${\mathsurround=0pt }#1\hfil##\hfil$\crcr
    #2\crcr\sim\crcr}}}

\documentstyle[12pt,aas2pp4]{article}

\slugcomment{Draft: \today}

\begin{document}

\title{A New High Resolution CO Map of the inner $2.'5$ of M51 \\}
\title{I. Streaming Motions and Spiral Structure}

\author{
S. Aalto\altaffilmark{1,2}
}

\author{ 
S. H\"uttemeister\altaffilmark{3,4}
}

\author{ 
N. Z. Scoville\altaffilmark{1}}

\and
\author{
P. Thaddeus\altaffilmark{3}}

\altaffiltext{1}{Division of Physics, Mathematics \& Astronomy, Caltech 105-24, 
Pasadena CA 91125}
\altaffiltext{2}{Onsala Observatory,
    S-439 92, Onsala, Sweden}
\altaffiltext{3}{Center for Astrophysics,
60 Garden Street, Cambridge MA 02138}
\altaffiltext{4}{Radioastronomisches Institut, Bonn University \\
Auf dem H\"ugel 71, D-53121 Bonn, Germany}

\renewcommand{\labelitemi}{  }

\begin{abstract}
The Owens Valley mm-Array has been used to map the CO 1--0
emission in the inner $2'.5$ of the grand design spiral galaxy M51 at $2''-3''$ resolution.
These new images reveal the molecular spiral arms with 
unprecedented clarity -- the emission in the two major arms (NE and SW)
originates from supermassive cloud complexes, Giant Molecular Associations (GMAs),
which are for the first time
resolved both along and perpendicular to the arms. The overall
morphology of the CO emission is symmetric in reflection
about the nucleus with major complexes 
occuring opposite each other in the two major arms.
On the other hand, the CO line flux in the area of the SW arm
closest to the nucleus is approximately twice as bright as that
from the analogous location in the NE arm. 

Streaming motions can be studied in detail and appear with great clarity along the
major and minor axes of M51. The streaming velocities are very large, 60-150 \kms. 
Our maps offer, for the first time, sufficient resolution to resolve the structure in the molecular
streaming motions. Both the radial and tangential velocity components show
steep gradients in qualitative accordance with predictions by the density wave models of Roberts and Stewart (1987).
Our data thus support the presence of galactic shocks in the arms of M51. 
In general, velocity gradients across arms are higher by a factor of 2-10 than previously found.
They vary in steepness
along the spiral arms, becoming particularly steep {\it in between} GMAs. 
The steep gradients cause conditions of strong reverse shear in several regions in the arms, and
thus {\it the notion that shear
is generally reduced by streaming motions in spiral arms will have to be modified}. Of the three GMAs
studied on the SW arm, only one shows reduced shear. 

We find an unusual structure, an expansion in the NE molecular arm at 25$''$ radius
(1.2 kpc) southeast of the center. This broadening occurs right after
the end of the NE arm at the Inner Lindblad Resonance (ILR). 
Multiple-peak spectra, velocity twists and structures with apparently
high velocity dispersion are associated with this feature.

Bifurcations in the molecular spiral arm structure, at a radius of $\approx$ 3.3 kpc (73$''$),
may be evidence of a secondary compression of the gas caused by, and
occuring near, the 4/1 ultraharmonic resonance. 

Several molecular spurs protrude from
the main spiral arms, in particular on the western side of the map. Inside the radius of
the ILR, we detect narrow ($\approx 5''$) molecular spiral arms possibly related to the K-band arms
found in the same region. We find evidence of non-circular motions in the inner 20$''$ of M51.
The magnitude of these deviations is 20-30 \kms, and they are consistent with gas
on elliptical orbits in a bar.

\end{abstract}


\section{Introduction}

The Whirlpool Galaxy M51 at 9.6 Mpc (Sandage and Tammann 1975) is one of the closest 
grand design spirals, rich in molecular gas, much of which is
found in the spectacular spiral arms (e.g. Young and Scoville 1983, Rydbeck \etal\ 1985,
Lord and Young 1990).
The striking spiral pattern of M51 has stimulated many observational studies.
Single dish CO studies of the molecular cloud
distribution have been undertaken by Young and Scoville (1983), Rydbeck \etal\ (1985), Lord and Young
(1990), Garc\'{\i}a-Burillo \etal\ (1993ab) and Nakai \etal\ (1994). Higher resolution aperture synthesis studies
of the molecular spiral arms have been done by Vogel, Kulkarni and Scoville (1988), Rand and Kulkarni (1990),
Tosaki \etal\ (1991), Adler \etal\ (1992) and Rand (1993a). Strong kinematic signatures of
density waves are velocity
discontinuities, or streaming motions, across the spiral arms (e.g. Roberts and Stewart 1987). The
unusually large streaming motions (60-150 \kms) of M51 imply a very strong density wave.
The high resolution studies resolve massive cloud complexes, 
termed Giant Molecular Associations (GMAs) by Vogel, Kulkarni and Scoville (1988) along the main spiral arms and have also found a few
apparently isolated complexes in the interarm region (e.g. Rand and Kulkarni 1990; Adler \etal\
1992; Rand 1993a). Possible differences in properties between arm and inter-arm GMAs have been discussed
(Rand 1993ab) with the conclusion that the arm-GMAs are probably gravitationally bound objects while
the interarm GMAs are probably unbound.
It is still unclear, however, how density waves may affect the stability of the gas in the arms in detail.
High resolution studies of streaming motions within, and in between, GMAs have been lacking and it has been
assumed that streaming does not change along the arm. The
idea that arms are relatively shear-free environments where the streaming motions help
cloud formation has been widely accepted.

Spiral arms are more readily studied in gas, compared to stars (e.g. Lubow \etal\ 1986; Lubow 1993).
Sensitive CO observations may therefore lead to a deeper understanding of the underlying mechanisms
behind spiral structure. Determining the location of resonances is particularly important since it
helps distinguish between various theories of spiral structure.
For instance, theories differ on whether spiral structure can extend beyond the 4/1 ultraharmonic
resonance on the outside (e.g. Contopoulos and Grosb$\o$l 1986; Artymowicz and Lubow 1992),
and whether it terminates before reaching the
Inner Lindblad Resonance (ILR) on the inside (e.g. Lubow 1993).

In this first paper we will focus the discussion on streaming motions and
spiral structure. Newly discovered structure in the streaming motions on the main arms
and the implications for GMA formation and survival are discussed in section (4.1-4.3).
The location of the 4/1 resonance
and its possible relation to structures in the interarm emission is discussed in 
section (4.4). Evidence for a molecular spiral in the inner kpc 
is presented in section (4.5) together with gas on non-circular orbits as indications of a
central bar. Adopted properties of M51 are listed in table 1. 

In paper II we will focus on the smaller scale structure of the molecular gas.
In particular, the boundedness of clouds and possible differences between on-arm,
interarm and central region gas will be investigated, also in relation to tracers
like H$\alpha$ and infrared.

\section{Observations}

Aperture synthesis CO mapping of M51 was carried out
with the Owens Valley Radio Observatory (OVRO) millimeter array between October 1995 and
May 1996. SIS receivers of the six 10.4 m telecopes of the array provided typical
system temperatures of 500 K. The quasars 1150+497 and 0917+449 were used for phase
calibration, and Uranus and Neptune for absolute flux calibration. 
Synthesized beams are $2.''9 \times 2.''1 $ (PA=-81$^{\circ}$) for robust (Briggs (1995)) weighting, 
and $3.''9 \times 3.''3 $ (PA=-43$^{\circ}$) for natural weighting. 

Natural weighting, with all points covered in the uv-plane treated alike, emphasizes shorter 
spacings, since the uv-tracks are more densely packed close to the origin. It gives the best
signal-to-noise ratio, but at the cost of poorer beam shape. Uniform weighting assigns identical weight 
to given cells in the uv-plane, thus maximizing the resolution at the cost of sensitivity  (e.g. Sramek
and Schwab, 1989). Because
of the limited uv-coverage common to mm interferometry, especially in mosaicing mode, this weighting
scheme often yields poor results. Robust weighting, which we applied instead to produce our most 
highly resolved images, improves the signal-to-noise ratio over uniform weighting while still giving
higher resolution than naturally weighted maps.

At wavelength $\lambda = 2.6$ mm and resolution $2.''5$, a brightness
temperature of $\Delta T_{\rm B}= 1$ K corresponds to 67 mJy beam$^{-1}$.
The digital correlator was centered at $V_{\rm LSR} = 460$ \kms, and provided
a total velocity coverage of  333 \kms, from 351 \kms\ to 601 \kms.
Data were binned at a resolution of 3 MHz, corresponding to a spectral resolution of
7.8 \kms. We observed a total of 19 positions (listed in table 2) in three different array configurations
(compact, low resolution, high resolution) --- the shortest baseline was 15m
and the longest 240m. The largest structure we could therefore image was 18$''$. The primary
beam is 65$''$ (FWHM) and the grid spacing of the 19 positions
was 30$''$, chosen to give approximately constant noise across the map.

We took 10 12-hr tracks per configuration for a total of 360 hours of telescope time. The observing scheme
was designed to minimize calibration inconsistencies within the final map: All positions were observed
sequentially, with only 2 minutes spent on each field at a time. Thus, each of the 19 positions was
observed 10-15 times during each track, giving a very homogeneous final data set.
Individual maps were made and deconvolved with the NRAO AIPS and the Caltech DIFMAP (Shepherd \etal\ 1995) 
software packages. The AIPS LTESS task, a
mosaicing procedure that corrects for primary beam attenuation, was used to combine all 19 positions into
a single datacube. The rms noise is 19 mJy per channel (0.2 K) throughout the whole naturally weighted map
(apart from the outer $\sim$ 3$''$ before cutoff, where the noise reaches 35 mJy per channel).

\section{Results}

\subsection{Naturally weighted intensity map}

Figure 1a shows the naturally weighted integrated intensity map
with $3.''5$ resolution. The molecular spiral arms appear with great clarity, and previously unknown 
structures on, and between, the main arms are revealed. A comparison with optical
pictures shows that the main CO arms trace the dust lanes beautifully. The map is dominated by
two main spiral arms of which the SW (M1) arm is the brightest, its flux being greater by a factor
of two than that of the NE arm (M2). The M1 arm is also broader than the M2 arm by $\approx$50\%.
The CO emission originates mostly from large cloud complexes (Giant Molecular
Associations, GMAs).
The brightest ($T_{\rm B}({\rm CO})=8$ K) GMA on the
M1 arm has a FWHM size of 11$''$ by 6$''$ (512 by 280 pc) and sits on a ridge of CO emission
20$'' \times 7''$ in size. Along the SW M1 arm we count 16 GMAs, with
sizes varying from 6$'' \times 3.5''$ to 14$'' \times 8''$. Local, distinct peaks in the intensity distribution were
counted as GMAs. Sizes for these peaks were determined using the 
AIPS IMFIT routine. Extended emission associated with a peak (at a level 20\% or less
of the peak) was not counted. The real number of GMAs in M1 is thus likely
larger than 16. 

A more thorough analysis of the size and structure of the cloud associations will be given in
paper II. Clouds previously studied by Rand (1993) (A4, A8 and A9) belong to the SW M1 arm and
are indicated in figure 1a. The first gaps in intensity along the main arms occur symmetrically for
both at a radius of 55$''$ where, as figure 1a shows, the arms first cross the major axis. 
The NE spiral arm M2 can be traced inwards to a galactocentric radius of 28$''$ (1.3 kpc), 
where the CO contours and the associated optical dust lane make a sharp turn eastward and the chain
of GMAs ends. The first bright GMA on the SW M1 arm appears at a radius of about 22$''$
(1.0 kpc) where its lowest contours bend eastward towards the nucleus. At lower radii the cloud associations
become poorly defined and fainter.
Feature C (in figure 1a) is a somewhat peculiar structure, a broad expansion 
perpendicular to the molecular arm southeast of the center. There is a similar, but less spectacular,
feature on the western side of the nucleus. Comparison of an optical image of M51 with our CO map shows
that the pattern of dust lanes spreads in a similar fashion to the CO emission, and that inside the C region
the dust lane structure is disordered. Feature C occurs right after the ending of the main
M2 arm at the claimed position of the ILR (Tully 1974b). 
We detect hints of molecular spurs, the features 
S1, S2, S3, S4, S5 in figure 1a, 
protruding at an angle (45$^{\circ}$ - 90$^{\circ}$) from the main spiral arms mostly on the western
side of the galaxy.

A fairly recently discovered feature in M51, also apparent in our map, is the compact molecular gas
concentration
at the nucleus of the galaxy, first detected in HCN by Kuno \etal\ and mapped at a
higher resolution of 1$''$ in CO 2--1 by Scoville \etal\ (1997).

There is evidence in the present study of bifurcations in the  molecular
spiral arm structure --- possibly as a result of the 4/1 ultraharmonic resonance
caused by the spiral potential (e.g. Shu, Milione and Roberts (1973); Artymowicz and Lubow
1992 (AL)). Two faint chains of molecular clouds, 
B1 and B2 split off from the main spiral arms at a radius of roughly 3.3 kpc (71$''$). 
Structures previously classified as discrete interarm clouds (I1,I2 in Rand 1993a) are part
of the B1 chain of clouds.

\subsubsection{The reality of the B1 feature}
   
The constant offset between the main arm M1 and the secondary arm B1 raises
the concern that B1 is spurious, a sidelobe artifact of M1, possibly the
result of incomplete cleaning or miscalibration. S. Vogel (private
communication) raised this possibility because a new CO map of M51 made with
the BIMA array does not show B1, but when he simulated observations of the
BIMA map using some (but not all of the OVRO configurations), an arm-like
feature appeared that is similar to if not identical to B1. In view of this
possibility, we carefully
reexamined evidence for and against the reality of the B1 arm. We conclude
that it is probably real for the following reasons:

i) The brighter emission peaks along B1 are clearly evident in
IRAM and OSO single dish CO maps (e.g. Garcia Burillo \etal\ 1993a
(fig 1), Rydbeck \etal\ 1991). The HI map by Rots \etal\ (1990) shows
structure that is coincident with B1. A number of faint H$\alpha$ 
emission regions are also seen along B1 (Vogel \etal\ 1993).

ii) A radial position-velocity cut through M1 and B1 reveals
that although there is some overlap of the velocity ranges 
of emission between M1 and B1, the peak velocity of B1 
is redshifted by 20 \kms\ relative to M1.  The velocity width of B1
is also narrower by a factor of two (see figure 1c). It is significant too that, on the
assumption of a flat rotation curve, B1 is at the expected velocity for its
radius. Lastly, we note that the streaming velocity structure from the main
arm to the bifurcated material is continuous at the beginning of B1.

iii) The new BIMA M51 CO map (Helfer \etal\ 1999, in preparation) 
does in fact show several low signal-to-noise features near the peaks in B1. 
Since the BIMA map is of lower sensitivity and resolution than ours, we think
that the two data sets may not be inconsistent. 

iv) While Vogel's simulation of the OVRO map does produce 
false arms, these do not agree very well in structure and 
position with our respective arms B1 and B2. The B1 feature in the
simulation is shifted 5-10$''$ inward from our B1, and is straight, not curved.
This last argument is of course not conclusive, since
the simulation did not include the OVRO compact configuration (which
probably would reduce the spurious features)
but in any case, if the residual sidelobes were the result of
miscalibration, the simulation would never exactly 
reproduce the artifacts. 


v)  The secondary structures survive deep cleaning, with various
approaches such as varying gain and number of iterations, as well 
as the position and number of cleaning boxes. They are also visible 
in a robustly weighted map, where the sidelobes are less 
prominent. The B features are also apparent in a map not biased 
toward positive emission. 

vi) The B1 feature is well inside the primary beam half 
power points (see table 2 for pointing positions) which means 
that it is not an edge effect of the map. The outer cutoff 
of the naturally weighted map (the outer dotted line in figure 1a) 
is somewhat (about 5$''$) inside the half power points of the 
outer pointings to show a map with (almost) constant noise. 
Note that the outer sensitivity cutoff of the robustly weighted map (figure 1b) occurs
at lower radius than for the naturally weighted one. 

Lastly, we note that the possibility that the B features are aliasing effects is
excluded because the size of the FFT was four times the map size.

Between B1 and the bright M1 arm, there is a shorter chain of interarm clouds (feature D, indicated by a
dotted line in Fig. 1a). These features also appear to be real structures for the same reasons
as above.

\subsubsection{Radial flux distribution}

The total integrated flux in the whole OVRO map is $2.7 \times 10^3$ Jy \kms. We have used the OSO 20m
single dish map of Rydbeck \etal\ (1985) to compare the total CO emission with that observed here 
interferometically. Inside a radius of 30$''$ the flux recovered by the OVRO array is 810 Jy \kms\
while the OSO single dish finds 1600 Jy \kms. Inside a radius of 60$''$
the OVRO flux is 1700 Jy \kms\ and the corresponding value for OSO
is 2800 Jy \kms, so that the OVRO array recovers  50-60 \%
of the OSO single dish flux in the region studied. Comparing our map with the NRO single dish map
of Kuno \etal\ (1995), we recover 30\% of their single dish flux in the inner 25$''$.  One large error source is
thus the relative calibration among telescopes, with OSO and NRO obviously differing by almost a factor of two.
Adler \etal\ (1992) claim to recover 50\% of the total OSO single dish flux in their BIMA aperture
synthesis CO map.

Figure 2a shows the brightness of each radial bin, the OVRO mean radial CO distribution from
the naturally weighted map. 
For the adopted, radially independent conversion factor (see table 1) between CO luminosity and H$_2$ mass, this
figure shows the change in the mean gas surface density with radius. 
The gas surface density increases steeply beyond a radius of roughly 8$''$ and reaches a
maxium (166 M$_{\odot}$ pc$^{-2}$) at a radius of 28$''$ (1.3 kpc) which is that of the
bright peak in the SW arm. There is then a sharp drop to a minimum at $R=52''$ (2.4 kpc),
reflecting the first intensity gap in the main spiral arms. The brightness then
increases somewhat again, and after 90$''$ it falls off steeply to the edge of the map at 95$''$.
From 28$''$ to 70$''$, the average gas surface density drops by a factor of three.
Kuno \etal\ (1995), from the NRO single dish map,
find a drop by a factor of $\approx$4 in the same interval, but their radial curve is less sharply peaked,
either because of their poorer resolution, or because the flux
undetected at OVRO is more smoothly distributed. The single-dish gas surface density overall may
drop somewhat faster with radius than that detected here, so it is possible that the interferometer
recovers a larger fraction of the single-dish flux with increasing radius --- consistent
with an increasing arm-to-interarm ratio with radius.

Figure 2b shows the variations of the CO intensity along the two major arms. The resolution has been smeared
to 10$''$. The clumpy nature of the arms is evident, as well as there being a striking regularity of
the distance, some 30$''$ - 40$''$, between the major concentrations of gas along the arms. These
peaks even occur at (roughly) the same spiral phase on both arms. These symmetries were also noted by
Garc\'{\i}a-Burillo \etal\ 1993a. A similar effect in the optical was found by Elmegreen, Elmegreen
and Seiden (1989) who interpreted it as an intrinsic part of the stellar wave pattern.

\subsection{Robustly weighted intensity map}

The robustly weighted map (figure 1b) recovers about 73\% of the flux of the naturally weighted
map in the same region, and much of the faint interarm structure is lost, although the brighter
parts of the B1 and B2 features are present. The main arms
are (not surprisingly) narrower in this map and the spurs are not
as clearly visible. In the inner arcminute, less flux is lost (the map recovers about 85\% of
the naturally weighted flux) and structures that could be narrow spiral arms can be discerned.

\subsection{Comparison with previous aperture synthesis maps}

Because of the high sensitivity, high resolution and improved uv coverage in our map, we find structures
undetected by previous studies. These structures include most of the spurs, the C feature,
parts of the B and D features, and the narrow molecular arms in the inner arcminute.

Our maps in general agree well with those obtained at OVRO by Rand and Kulkarni (1990, RK)
at 8$''$ resolution. There are hints in their data of the S1 spur and the B2 bifurcation; 
the positions of GMAs and emission gaps also agree with ours. There is some disagreement
between the RK map and ours in the intensity distribution across the map; the SW arm is too faint
in their map and Rand (1993, R93) also points out that that part of their map is suspect owing to 
calibration errors. R93 mapped two overlapping 1$'$ fields that cover part of the M1
arm, and some interarm emission, with the OVRO array at $3.''5$ resolution; his arm structure and
cloud positions agree well with ours, although he misses much structure which we see (e.g. the
A10 cloud) owing to his poorer uv coverage. Rand found two interarm emission features (I1 and I2 in
R93) that we  assign to the B1 cloud chain.
Adler \etal\ (1992) presented observations of M51 obtained with the Berkeley-Illinois-Maryland Association
(BIMA) interferometer; they observed two $2'$ fields with a synthesized beam of $7'' \times 11''$. 
Since their shortest
baseline (8 m) was about half of ours, they were sensitive to larger structures. Given the much larger
synthesised beam of the Adler \etal\ map, the qualitative agreement
between their map and ours is as good as can be expected.

\subsection{Velocity Field and Velocity Dispersion}

The naturally weighted velocity field is shown in Figure 3 [Plate 1]. Velocities range from 350
(deep blue)  to 550 (red) \kms, with the gas approaching us in the northern part of the galaxy and
receding in the south. Streaming motions appear very
clearly in the velocity field as jumps and discontinuities in the contours across spiral
arms, in particular along the SW main arm, M1. Radial streaming motions, best seen along
the minor axis, are prominent in the M1 arm (west)
while none are readily apparent in the east. Streaming motions along the major
axis are more symmetrical with respect to the M1 and M2 arms.

Because of the significantly increased resolution of the robustly weighted map, streaming motions on
the main arms are better resolved than in the naturally weighted map, in particular where the M1
arm crosses the major axis. Sections of the robustly weighted velocity field are displayed in figures
6a and 6b.

The velocity contours associated with the C feature are somewhat twisted, indicating
a complex velocity field in this region. Spurs S1 and S4 show velocity reversals
as they stretch out into the interarm region, S2 and S3 show strong velocity gradients
but no reversals, while S5 shows no velocity structure
at all. Along S4 the velocity shifts from 500 to 540 \kms\ and then back to 500 \kms\ as we move into 
the arm. Within S1 the shift is  from 400 \kms\ to 450 \kms\ and then back again. The red feature in
the center of the map is the small rotating nuclear gas disk which, as 
previously noted by Scoville \etal\ (1997), is redshifted with respect to 
the systemic velocity by $\approx 40$ \kms.

The naturally weighted dispersion map is shown in figure 4 [Plate 2]. The map shows
the one dimensional dispersion ranging from $\sigma_{\rm 1d}=5 - 20$ \kms.
Only one value of $\sigma_{\rm 1d}$ is fit, even if the spectrum in question consists
of multiple peaks. The red regions in the map indicate high dispersion, $\sigma_{\rm 1d} \gapprox 18$
\kms. Most of the high $\sigma_{\rm 1d}$ structures may be caused by unresolved streaming motions
(e.g. where the western minor axis cuts the M1 and B1 arms). Positions of strong streaming
motions, indicated in the velocity field, do indeed correlate with regions of high dispersion. 

Other regions of disturbed velocity contours, such as the C region, and the bifurcation areas also show
spots of high dispersion gas. In the C-region, several of these high dispersion spots
are caused by multiple peaked spectra, but there is a small area of faint, broad line emission
at the southern tip of the C feature, right before the feature narrows and $\sigma_{\rm 1d}$ drops to
6-10 \kms. It is evident too that the nucleus is a region of high velocity dispersion.

Measured on the naturally weighted map, the average velocity dispersion in the M1 arm is
$<\sigma_{\rm 1d}>=12 $ \kms\ and 
$7 $ \kms\ in the M2 arm before the emission gap at radius 55$''$. After the gap, the
average dispersion is about the same for both arms, $<\sigma_{\rm 1d}>=9 $ \kms.
The average dispersion on the B1 structure is $<\sigma_{\rm 1d}>=6 $ \kms.
The emission from the molecular spurs and the D clouds is in general of low dispersion (4-8 \kms), except for 
a few positions along S1.
The general result, that the dispersion of the interarm emission gas is lower than that of the arm
emission, is consistent with the single dish result (at lower resolution) by Kuno and Nakai (1997) who
suggest that the dispersion is higher on the arms because of
the streaming motions. Indeed, $<\sigma_{\rm 1d}>$ on the M1 arm is reduced from 12 to 9 \kms\ at the
higher resolution offered by the robustly weighted map.

\subsection{Molecular mass}

We estimate the molecular mass in the inner $2.'5$ to be $1.9 \times 10^9$ M$_{\odot}$, by adopting
a standard CO to H$_2$ mass conversion factor (see table 1).  Since between 70\% and 40\% of the total
flux is not detected by the interferometer, the real molecular mass in this region could
be as high as $6 \times 10^9$ M$_{\odot}$ assuming that the same mass conversion factor holds
for the arm and interarm gas.
However, according to Nakai and Kuno (1995), Arimoto, Sofue and Tsujomoto (1996) and Adler \etal\
(1992), the CO to H$_2$ conversion factor for the inner 4.5\,kpc of M51 is
lower by a factor of 2--3 than the Galactic mean value adopted here, increasing with radius. These 
values are based on CO-independent determinations of molecular
masses by applying optical extinction measurements or assuming virial equilibrium and comparable to the 
lowest values suggested for our Galactic disk.

\subsection{Arm-interarm contrast}

Using the naturally weigthted map (figure 1a), we partitioned the observed emission into arm and
interarm regions to determine the interferometric arm to interarm ratio. The arms were defined by
the lowest contour in figure 1a.
We have classified everything outside the main stellar arms as
interarm --- for instance, the B1 and B2 features have been considered interarm
as have the spurs protruding from the main arms. In Jy/beam \kms\ the mean flux in the arms
is 3000 and that in the interarm regions is 190 resulting in an
average arm-to-interarm ratio of about 16.
The interarm regions have a total integrated flux of 390 Jy \kms, and the arm regions
have 2300 Jy \kms. This gives a total flux ratio of 5.9, and with
the standard conversion factor (see table 1) the mass in the interarm 
molecular gas observed here is about $2.9 \times 10^8$ M$_{\odot}$.

The {\it actual} arm-to-interarm flux ratio depends on how the missing flux is
distributed. If the unseen emission is evenly distributed across the map, the actual arm-to-interarm
ratio will be lower than that just derived. The total flux in our map is 2700 Jy \kms; assuming 
an equal amount is unobserved, and evenly distributed across the map, the arm-to-interarm
contrast is reduced to about 3. The arm to interarm mass ratio would then be 1.3 instead of 6. 
In the same way, Vogel, Kulkarni and Scoville (1988) estimate from their interferometric CO map
an arm-to-interarm mean flux ratio of 2.4-3.0. However, Adler \etal\  (1992) argue that when
combining a single dish map (the OSO 20m map of Rydbeck \etal\ 1985) with their aperture synthesis
BIMA map, most of the missing flux is instead distributed in close association with the arms, making them
broader, and they measure an arm-to-interarm mean flux ratio $\gapprox$4.6.

Single-dish measurements detect all of the flux, but may lack sufficient resolution to clearly distinguish
between arm and interarm regions. Previously measured single-dish CO arm-to-interarm ratios for M51
show average values of the mean arm-to-interarm flux ratio of 3-5, but also significant variation with radius:
Garc\'{\i}a-Burillo \etal\ (1993a) measure a {\it peak} arm-to-interarm ratio that increases with radius
from 2 to 6. Kuno and Nakai (1997) find very similar values of 1.5-7, again increasing with radius,
with the NRO telescope.

The molecular arm-to-interarm flux ratio is interesting as an indication of the presence and
strength of a density wave in a galaxy (e.g. Roberts and Stewart 1987; Elmegreen 1988). 
The density contrast increases with increasing strength of the spiral gravitational field (e.g.
Shu, Milione and Roberts (1973)). Roberts (1993), for example, finds an arm-to-interarm contrast
of 6 for a 10\% modal spiral driving. While our map is the first large scale interferometric data set to 
detect significant interarm emission in M51, the contrast of 12 we find is still likely to be an upper limit.
Thus, the arm/interarm contrast from our data cannot be used to constrain spiral density wave models 
beyond stating that it is generally compatible with them. 
Since the ISM passes into
and out of the arms cyclically, the flow of matter must obey the continuity
equation in order to conserve mass. Thus, if $t_{\rm arm}$ is the 
time a molecule stays in the arm and $t_{\rm inter}$ is the time it spends in the interarm
region, then ${M_{\rm arm} \over t_{\rm arm}} = {M_{\rm inter} \over t_{\rm inter}}$. In principle, it
should be possible to determine arm residence timescales from this requirement. Our limits of
arm-to-interarm mass ratio of 1.3 to 5 are not restrictive enough to pursue this,
but a combination with single dish data might make a detailed analysis feasible in the future.

\section{Discussion}

\subsection{Streaming motions}

M51 is one of the galaxies where the strongest evidence for density waves has been
found (e.g. Rand 1993b). The magnitude of the streaming motions is unusually high 
(e.g. Tully 1974b; Rydbeck \etal\ 1985; RK; Vogel \etal\ 1993) amounting to about 100 \kms\ in the
plane of the galaxy --- a factor
of two higher than predicted by linear theory, or non-linear simulations (e.g. Tully
1974b; Roberts and Stewart 1987). In previous CO work at OVRO
(Vogel, Kulkarni and Scoville 1988; RK; Rand 1993a (R93)) both radial and tangential streaming motions in the
arms of M51 were detected, and
streaming motions are also evident in single dish CO data (e.g. Rydbeck \etal\ 1985;
Garc\'{\i}a-Burillo \etal\ 1993ab; Kuno \etal\ 1995).

Density waves are expected to cause velocity jumps and discontinuities across molecular spiral arms.
Tangential streaming is expected to be the most apparent along the major axis, radial streaming
along the minor axis. Along the major axis the highest local velocities (negative
on the blueshifted side of the galaxy and positive on the redshifted side) are found on the 
outside of the arm, and on the inside one finds the lowest local velocities. Outside
of corotation, the situation is reversed. On the minor axis, a radial velocity discontinuity, a decrease
followed by an increase, across the arm is expected. In our map the density wave perturbations begin
(for the M1 arm) at a radius of 22$''$ ($\approx$ 1 kpc), and it appears
that the streaming motions are more distinct in the brighter, broader M1 arm than in the
M2 arm.

Roberts and Stewart (1987) (RS) have made simulations of the response of the interstellar medium
to a spiral potential, treating individual molecular clouds as point particles. In their model the 
spiral forcing amplitude relative to the central axisymmetric force is moderate, i.e. 10\%.
They studied models that included dissipative cloud-cloud collisions and simplified cases without
cloud-cloud collisions.  They find that dissipative cloud-cloud collisions are
necessary to stabilize the streaming motion pattern in time, and will produce stronger, sharper velocity
and density profiles than the non-collisional case. Figure 5 shows the cloud density, tangential and
radial velocity components plotted vs. spiral phase, from RS dissipative case simulations.

Note that, for M51, the notion of a (quasi) steady density wave (e.g. Lin and Shu 1964) may be
incorrect if the waves are triggered by swing amplification by the passing of the companion galaxy
(e.g. Toomre 1981). In this case, the density wave pattern is transient and a steady streaming motion
pattern may be unlikely.

\subsection{Tangential streaming}

According to the RS model (figure 5) the tangential streaming 
decreases as gas approaches an arm and then quickly increases across the arm.
There is a very sudden change in slope, a sharply pointed trough, indicating
a shock. 

Where both the M1 and M2 arms cross the major axis (at radius 55$''$) there is a gap in CO intensity.
We have therefore studied the streaming motions in these arms 
at $\pm 10^{\circ}$ off the major axis.
Figure 6a and 6b show the robustly weighted ($2.''5$ resolution) velocity
field (contours) overlayed on the integrated intensity (grayscale) in the southern and
northern arm-crossing regions, respectively. 
In figure 6a the GMAs previously studied by R93, A8 and A9, are marked as well 
as a GMA between these that we call A10. {\it The non-linearity of the streaming motions is
clearly evident here: The velocity contours are not evenly spaced across the arm, but
instead a narrow region of densely spaced contours indicates a very steep increase in velocity}.

Furthermore, figure 6a clearly shows that streaming is highly uneven along the arm. 
Most of the streaming occurs {\it before} cloud A9 and {\it through} A10. In A8 some streaming
can be detected but the gradient is not as steep as at A10. Streaming
velocity gradients through five GMAs and three inter-cloud
regions (IC1,IC2,IC3 in figues 6a and 6b) are listed in table 3. As is evident from the table, gradients
vary significantly from GMA to GMA, the highest being that through A10, with 700 \kms\ kpc$^{-1}$ and
the lowest through A9 with 65 \kms\ kpc$^{-1}$. In general, {\it gradients appear steeper in between GMAs}.
IC1 is the region upstream of A9 and between A9 and A10 and has the highest velocity gradients
measured in this map, 800-1300 \kms\ kpc$^{-1}$. Although Rand (R93) found a similar total magnitude of
the streaming motions, he measured a significantly lower gradient of $\approx 100$ \kms kpc$^{-1}$ at
the southern arm crossing. He also assumed the gradient to be constant along the arm. The difference
probably results from Rand's poorer uv coverage and angular resolution.

The highest total magnitude of the tangential streaming motions, 150 \kms, is measured through A10.
The magnitude of the streaming motions is somewhat lower on the NE arm, about 100 \kms\
across A2.

\subsubsection{Arm shear}

Why do we see such structure in streaming motions and streaming motion gradients along the arm? Is this
just an effect of the GMAs progressing across the arm, with A9 perhaps
having entered a region of low velocity gradient further downstream, or is it related to the cloud's
internal response to streaming? 
Indeed, the presence of GMAs and the structure of the streaming are related,
since the steepness of the gradient increases between the complexes. It seems
that the presence of clouds does indicate lower shear regions, either because clouds
form preferentially in such regions or because many of them have detached themselves
more or less completely from higher shear regions.
Streaming motions have been suggested
as conducive to molecular cloud formation. A local solid body rotation is supposedly 
created which increases shear time scales giving the gas enough time to form
complexes before they pass into the harsh inter-arm region.
The shear time-scale is the inverse of Oort's A-constant (e.g. Elmegreen 1988):

$$
T_{\rm shear}=A^{-1}=2 ({v_c \over R} - {dv_c \over dR})^{-1}_R
$$

where $R$ is the radius and $v_c$ the circular velocity. It is clear from this relation that for the shear to
be successfully counteracted, ${dv_c \over dR}$ must be very close to ${v_c \over R}$,
i.e. local conditions must approach solid body rotation. Rand (1993b) used this relation to estimate shear time
scales. For clouds on the NE (M2) and SW (M1) arms at 2.7 kpc from the nucleus 
values of ${dv_c \over dR}$ close to 80 \kms\ kpc$^{-1}$ would reduce shear effectively. He
found velocity gradients close to 60 \kms\ kpc$^{-1}$ for the NE arm and about 100 \kms\ kpc$^{-1}$
for the SW arm. For both arms, shear time scales were thus increased by about
a factor of 5 compared to the differential rotation case, with the shear becoming slightly reverse 
in the south. 

However, for the same regions, our conclusions are different. 
In general, the gradients that we measure are at least a factor of two higher than what Rand (1993b) finds for
the southern arm crossing, implying that a condition of strong reverse shear will exist. To judge from
the above relation, this reverse shear will be even greater than shear in the pure differential rotation
case. Shear timescales for selected GMAs and intercloud regions are listed in table 3, together with the
corresponding Oort constants. In the regions of extremely
steep gradients found in between GMAs and through A10, this effect becomes so pronounced that it results in
significantly shorter shear timescales than from a flat rotation.  Thus, {\it the conclusion that shear
is generally reduced by streaming motions in spiral arms must be reconsidered. Of the three GMAs
studied on the SW arm, only A9 shows reduced shear.} 
Thus, conditions close to solid body rotation do occur, but apparently not {\it throughout} the arms.
Future statistical studies of cloud stability on the arms will have to take the varying
degree of shear into account instead of generally assuming that shear is low. For more typical spirals
where the average streaming motions are smaller, the shear may 
still be generally reduced in the arms. Since the situation is likely to be complex, however,
detailed high resolution studies of the velocity fields of the 
arms in a number of galaxies are necessery to decide this question.

Elmegreen (1988) points out that shear may not be such an important factor in forming complexes when the
gas is highly dissipative, since the incident kinetic energy will be removed by processes such as
cloud-cloud collisions.
In this context, the variation in steepness of the velocity gradient along the arm may be interpreted as
the GMAs having shed some of the strong reverse shear induced by the density wave, which
still persists between the clouds. A10 may still be in the process of such a dissipation. It is quite
suggestive that the region of reverse shear exists {\it upstream} of A9, suggesting the possibility that A9
once passed through this high-shear region on its way downstream.

\subsubsection{Interarm streaming}

The existence of the S1 and S2 spurs makes it possible to study the velocity structure of gas in
the interarm region. Figure 6c shows a position-velocity cut (pV-diagram) through the S1 spur. The
agreement with the RS prediction is good; the tangential velocity decreases just before the gas enters the
arm and then increases steeply in the arm. The expected
peak in dispersion (Roberts 1993) right before the arm is also present
in our data. We can measure the magnitude of interarm tangential streaming in these spur regions, and
find that the gradient is $\approx 100$ \kms kpc$^{-1}$, in the opposite direction to the arm gradient
as qualitatively expected. This reversed streaming is also expected at the outside
of the arm.

There is a strong concentration
of gas 15$''$-20$''$ away from the M2 arm. This possible GMA seems to exist where the steep
reversed tangential streaming begins to flatten out.  Note that the dispersion in the gas also decreases
in this region. Perhaps the D clouds have formed in similar calm regions in spurs between the arms.

Single dish studies suggest arm widths of $20''-30''$ instead of the $5''-15''$ we find here. 
The structure along S1 indicates that the velocity reversals occur $7''-17''$ away from the arm center.
The broader arm widths thus imply that the single dish surveys are picking up emission from the
velocity reversal regions. The reversed streaming in these regions may be more
disruptive to GMAs than even the high shear regions in the arm since the amount of
dissipation will decrease with distance from the arm center. Tidal forces and flow
expansion are other disruptive agents that may tear GMAs apart in the velocity reversal regions (Rand 1993b).
Perhaps the CO emission here largely originates in diffuse unbound gas that the interferometer will
filter out, but the single dish instruments can detect.

\subsection{Radial streaming}

Figure 6d shows a pV-diagram along the minor axis which goes through the two main
spiral arms at a radius of 25$''$ to 35$''$. Figure 6e shows a close up of the velocity
discontinuity across the M1 arm. There is good agreement with the RS radial streaming model
(dashed line in figure 6e), with the
steep decrease in velocity, the density peak of the arm occuring at the velocity minima,
and then the increase in velocity on the outer side of the arm.  The magnitude
of the shift is roughly 95 \kms. High resolution
is needed to resolve the shifts and even at $2.''5$ resolution, the shift is still somewhat
unresolved, but an estimate is that it occurs on roughly 5$''$ (230 pc), which means that the
gradient is 430 \kms\ kpc$^{-1}$. Rand (1993b), however,
finds only a 60 \kms\ shift in the ionized gas (H$\alpha$) at the same arm crossing. The reason for this
discrepancy is not clear, but if there is a displacement between the molecular and ionized gas,
the H$\alpha$ might miss the region of steepest gradient. Extinction may also hamper the interpretation
of the H$\alpha$ emission.

The radial streaming motions are not as pronounced on the eastern (M2 arm) minor axis.
Although a drop in velocity can be detected on the inside of the arm, it is of smaller
magnitude, 50 \kms, than on the west side, and the expected 
rise in velocity on the outside of the arm is absent.
The interpretation is also complicated by the multiple velocity structure of the gas inside
the arm, at radius 0-23$''$. The figure reveals two emission regions, one peaking at
440 \kms, the other at 490 \kms, although they are at the same radius along the same 
cut through the minor axis.
Vogel \etal\ (1993) also note that, for their
Fabry-Perot H$\alpha$ data, the streaming for the M2 arm where it first crosses the minor
axis is weaker than what is measured for the M1 arm on the 
corresponding position. They attribute this to the more complex gas response (as indicated by dust
lane morphology) of the M2 arm close to the oval distortion.

Although streaming motions can be observed along the
M1 arm all the way between the minor and major axis arm crossings, very little streaming
can be detected on the M2 arm until the major axis crossing in the
north. The reason for this lack of streaming in the first quadrant of M2 is unclear.
It can hardly be attributed to complex velocity structure which is suggested to
explain the lack of streaming on the minor axis (see above). A possible explanation is that the M2
arm in this region is only half as thick as the M1 arm and may be too narrow
for a display of the full velocity range of the arm streaming. We would then
expect to see streaming in M2 in another tracer than CO, sampling a different region
of the arm. Vogel \etal\ (1993), however, note the lack of streaming in M2 in their
H$\alpha$ map as well. Perhaps the reason can be found in an assymmetry in the matter
distribution.


\subsubsection{Galactic shocks}

The shape of the radial and tangential velocity components in figure 5 shows sharp deceleration and
troughs, signifying the presence of a galactic shock. In the early models, self-gravity was not
included and Lubow \etal\ (1986) suggested that introducing self-gravity would
smooth density and velocity profiles, removing the shock structures. Roberts (1993) however reports
that when including self-gravity in the simulations the abrupt velocity structure remains, even 
though the density profile becomes somewhat smoother. 
Our maps offer, for the first time, sufficient resolution to resolve the structure in the molecular
streaming motions. As we have seen above, both the radial and tangential velocity components show
steep gradients (see figues 6a, 6b, and 6e) in accordance with predictions by the RS models. The
region IC1 (in figure 6a), for example, clearly delineates the shock structure formed on the southern
arm crossing of the major axis. {\it Our data thus support the presence of
shocks in the arms of M51}. 
The fact that the shock structure and magnitude varies along the arms
indicates that local effects enter to modify the effects of large-scale dynamics.

Single dish surveys (Garc\'{\i}a-Burillo 1993ab; Kuno and Nakai 1997) report a more gradual change of the
velocity vector across the arms. Kuno and
Nakai (1997) find that the deceleration of the radial velocity component continues for about 1 kpc
($20''$). Because of this smoothness, they conclude that galactic shocks do not occur in the arms of M51.
At their resolution ($16''$), however, the sharpness of the velocity profiles will be smeared,
since the gradient occurs on a scale of $1''-5''$ and we suggest insuffcient resolution in the single
dish studies is the main cause of the discrepancy.

\subsection{Ultraharmonic resonance}

Lubow (1993) and Bertin (1993) discuss the behaviour of the cold gas component at
resonances and how one may use observations of gas in galaxies to detect resonance
locations. Due to its dissipative nature, the gas behaviour at the resonances is 
expected to be quite different from that of the stars, masking the
location of the resonance or, in some cases, revealing it through conspicious features. 
Shu, Milione and Roberts (1973) suggested that a second compression of the interstellar gas,
associated with an ultraharmonic $n$=2 resonance, may generate secondary spiral features. 
Patsis \etal (1997) studied the response of a gaseous disk to a spiral perturbation,
and their models show arm bifurcations and secondary
arms between the main arms. They suggest that
these bifurcations are the sign of the 4/1 ultraharmonic resonance and could serve
as good resonance indicators. Artymowicz and Lubow (1992, AL) 
also predict the launching of a second wave in the gas at the location of the lowest order
ultraharmonic resonance. The resonance may result in the termination, or severe depression,
of the main arm (e.g. Contopoulos and Gr$\o$sbol 1986; AL). Contopoulos and
Gr$\o$sbol also expect larger deviations from circular orbits near the resonance which
should result in increased radial streaming motions in the gas.

The arm bifurcations in our figure 1a are symmetrically located on
opposite sides of the galactic center at a radius of $\approx 71''$ (3.3 kpc at $D$=9.6 Mpc). 
They are very similar to those in the models by Patsis and coworkers, which leads us to suggest
that the B1 and B2 cloud chains may be faint, patchy, secondary spiral arms caused by the 4/1
resonance. According to Patsis \etal\ the actual resonance occurs somewhere
between the arm bifurcation and the termination of the arm. Rand (1993b) places the
4/1 resonance in M51 close to 3.9 kpc based on an intensity minima in the main CO arm.
He also finds large streaming motions (125 \kms) right before the minima
at 85$''$ radius. We find similar velocity shifts, 130 \kms, in the gas at the location of the two
bifurcations (see figure 3).


If the rotation curve is known, the location of one resonance will predict the position of the
other resonances (e.g. Binney and Tremaine, 1987). There are, for example, several studies that have aimed
at identifying the
location of the corotation resonance (Tully 1974; EES; Vogel \etal (1993); Garc\'{\i}a-Burillo 1993),
see table 4, and we can use it to predict the expected location of the 4/1 resonance. We adopt
$R$=6 kpc as the position of the corotation resonance, although the total range in table 4 is
5.8-7.4 kpc.
For a Mestel disk (i.e. a disk in which surface density is inversely proportional
to radius resulting in a flat rotation curve, e.g. Binney and Tremaine (1987)),  having corotation at
6 kpc will place the 4/1 resonance roughly at 3.8 kpc. 
The estimated positions for the ILR lies between 1.4-1.7 kpc (Tully 1974b; Garc\'{\i}a-Burillo \etal\ 1993,
see table 4), and for a Mesteldisk this places the 4/1 resonance between 3.0 and 3.9 kpc. EES suggests
that the Outer Lindblad Resonance (OLR) lies at $R$=8 kpc which, for the same model, places the
4/1 resonance at $R$=3 kpc.

\subsubsection{Two spiral patterns}

According to Contopoulos and Gr$\o$sbol (1986) the 4/1 resonance should represent
the limiting radius for (strong) spiral arm structure, and the main arms should stop not far
beyond. AL agree that the 4/1 resonance may result in strong losses, but the amplitude further
out at corotation could still be large.

In M51, the spiral pattern clearly continues beyond the suggested location of the 4/1 resonance. The CO arm
is indeed suppressed close to the resonance (e.g. Rand 1993b, see above) but
commences again at larger radius. It has been suggested that
there are two pattern speeds in M51, with the outer arms belonging to a different pattern, driven by
the companion galaxy NGC~5195 (Tully 1974b; Elmegreen, Elmegreen, and Seiden 1989 (EES); Vogel \etal\ 1993). 
EES, for example, propose 1. An inner spiral mode that has an Outer Lindblad Resonance (OLR) at the position
of a prominent arm intensity gap at $R$=8 kpc (pattern speed 90 \kms), 2. and an outer material arm that
corotates with the companion (pattern speed 22 \kms). They argue that the position of the ILR of the outer
arms occurs at the same radius as the corotation resonance of the inner spiral, and they suggest that
the outer spiral triggers a response in the vicinity of its ILR which stimulates corotation of an inner
spiral wavemode. The radius of the ILR of the outer spiral mode represents the inner limit to its extent
and therefore, the inner spiral structure extends beyond its 4/1 resonance, at least out to corotation.

Toomre (1981) argues that the grand design spiral structure of M51 is
largely the result of a strong, transient perturbation, and that the
density waves excited by it are likely to differ significantly from those
considered in the density wave theory of isolated systems. Specifically, 
the 4/1 resonance should not be excited in such a transient
encounter. As we have discussed , however,
the arm bifurcations, CO intensity minima and large streaming motions are expected signs of
the 4/1 resonance. EES also discuss optical evidence for the presence of a 4/1 resonance.
Further studies, theoretical and observational, are necessary to resolve this issue.

\subsection{Molecular structure in the central arcminute}

\subsubsection{Spiral structure}

The Inner Lindblad Resonance (ILR) occurs at the apparent termination of spiral structure
at a radius of $25''-30''$ (e.g. Tully 1974b; this work). However, narrow tongue-like features in the
central 25$''$ of our robustly weighted CO map (figure 7) hint at the
possibility of an inner compact pattern in M51. The possibility of such structure
was first raised by Zaritsky, Rix and Rieke (1993, ZRR), who from a K-band image residual
found evidence of spiral arms inside the ILR. They claim that these arms continue to wind
through an additional 540$^{\circ}$ beyond the ILR, ending at about
10$''$ from the nucleus. They further suggest the existence in the innermost 10$''$ of a small
bar, possibly an oval bulge. It is surprising to find spiral arms inside the ILR since
it is generally claimed that a density wave will not propagate there (e.g. Lubow 1993). By analogy with
the outer material arms, however, the structure within the ILR may not be part
of the same pattern as the spiral structure beyond a radius of 25$''$: arms, for
example, may be driven by a central bar, or by the spiral pattern further out. It is also possible
that the assumed position of the ILR is wrong and should be placed at much shorter
radius. This is a possibility since it is difficult to measure the rotation curve in the 
very center of the galaxy.
We have attempted to compare the ZRR K-band image of the inner spiral structure with
our robustly weighted CO map to see if the spiral arms in the ZRR map (dashed lines in
figure 7) can be identified with
the features in the molecular gas. Some of the narrow CO features in the inner arcminute
are found to agree in position with the K-band arms, but some do not. 
The width of these tentative molecular
arms is about 5$''$ and the structure is patchy, not tracing the arms fully, as is also
true for the K-band arms.

\subsubsection{The central bar}

The presence of a  bar in the inner region of M51 has been discussed by several authors
(e.g. Tully 1974b; Smith \etal\ 1990; Pierce 1986; ZRR; Garc\'{\i}a-Burillo \etal\ 1993b; Kohno \etal\ 1997). 
Tully (1974) identifies the ILR with the observed terminal radius of the (main) spiral structure
at 25$''$ and from deviations of circular motions in his H$\alpha$ data suggests the existence of an
oval dispersion ring inside the ILR with PA 130$^{\circ}$, 
causing gas streaming on elliptical orbits, as found by Tully (1974b). There seems to be a general consensus that
the PA of the bar, if it exists, is approximately 130$^{\circ}$, but the length of the bar is fairly
uncertain: between 10$''$ and 20$''$ (e.g. Pierce 1986; ZRR).

Our data show no continuous gas structure that fills the central bar. This may be because the
bar is filled with diffuse faint molecular gas that the interferometer will filter out, or simply
because the bar is deficient in molecular gas, or it may mean there is
no central bar at all. The robustly weighted map does show some faint emission connecting the
central gas structure to the emission region to the northwest (see fig 1b) but this structure is
only a few arcseconds long; discussion of it will be left to paper II.

In order to look for evidence of non-circular motions in the center, we used the AIPS GAL task to
subtract a rotation curve from the naturally weighted velocity field. Initial values were adopted from
Tully (1974a): PA=$170^{\circ}$,
$i=20^{\circ}$, $v_{\rm sys}$=464 \kms. We adopted a rotation curve of the form
${v \over v_{\rm max}}=1 - e^{-ln(100 {R \over R_{\rm max}})}$. The resulting central
residual map is shown in figure 8. 
In the inner 30$''$ there are two main deviations from the general rotation 
curve, a blueshifted region to the east and a redshifted region to the west.
The size of the region of deviating velocities is roughly $20'' \times 10''$ and the PA
of the joined structures is about 125$^{\circ}$. 
The magnitude of these deviations is 20-30 \kms\ and they are consistent with gas on elliptical
orbits in a bar. The region of large residual velocities agrees roughly with
the end of the K-band spiral arms of ZRR --- and with the beginning of the 
bar they suggest is present in the inner $10''-20''$.

\subsubsection{The nature of structure C}

The structure we have named the C feature (figure 1a) is located at the inner end of the main NE (M2)
spiral arm, at what we agree is the radius of the ILR of M51. The feature also lies
at the eastern end of the possible central bar, although we note that the corresponding feature
at the western end is not so spectacular.  Tully points out that the regions where the spiral arms
meet the material on elliptical
orbits (the ``dispersion ring'') usually show evidence of abrupt velocity discontinuities, pronounced dust
lanes and strong radio continuum emission.  
In our data there are multiple spectral peaks at several positions in this
region. As we can see in the dispersion map (figure 4), the very inner region of
the C feature has quite narrow lines (low dispersion) while the more complex velocity
structure occurs around the ``edges'' of the feature (as indicated by red regions in
figure 4). 
In the lower (southern) region there are multiple spectral featues, where the less prominent feature
is at velocities 20-50 \kms\ lower than the main feature (figure 9a). It is possible that the
low-velocity-feature (which
is about 3 times fainter than the main one) originates in gas on bar orbits while
the brighter emission comes from gas on circular orbits.
The upper (northern) ``red'' region also consists of double spectral components, but here
the high velocity feature is the fainter one (figure 9b). 
There are also some positions where the high dispersion is caused by broad lines, $>$ 50 \kms,
rather than multiple spectra.
We speculate that the disturbances in feature C are the result of high velocity collisions between
molecular clouds in different orbits. If cloud-cloud collisions occur anywhere, they are most plausible
in such a region,  where central spiral arms,  main spiral arms, and gas
on elliptical orbits converge.

\section{Conclusions}

\noindent
We have undertaken a new sensitive, high resolution CO study of the inner $2.'5$ of
M51. The main results are:

\begin{itemize}

\item{} The emission in the two major arms (NE and SW)
originates mostly from supermassive cloud complexes (Giant Molecular Associations, GMAs)
which are for the first time
resolved both along and perpendicular to the arms. 
The brightest ($T_{\rm B}({\rm CO})=8$ K) GMA on the
SW arm has a FWHM size of 11$''$ by 6$''$ (512 by 280 pc) and sits on a ridge of CO emission
20$'' \times 7''$ in size. Along the SW arm we count 16 GMAs, with
sizes varying from 6$'' \times 3.5''$ to 14$'' \times 8''$. 
The overall morphology of the CO emission is symmetric in reflection
about the nucleus with major complexes occuring opposite each other in the two major arms.
The CO line flux in the inner area of the SW arm
is approximately twice as bright as that
from the analogous location in the NE arm. 
The typical one-dimensional velocity dispersion is 7-15 \kms\ on the arms 
and the arm width varies between 5$''$-15$''$. The SW is broader than the NE arm by $\approx$50\%.

\item{} The average aperture synthesis arm-to-interarm CO flux ratio is
about 12. We detect five times more flux in the arms than in the interarm region. The total flux
detected by the interferometer is $2.7 \times 10^3$ Jy \kms, which is about 50\% of the Onsala 20m
single dish flux in the same region. For a standard Galactic CO mass calibration factor (see
table 1) the molecular mass seen by the interferometer is $1.9 \times 10^9$ M$_{\odot}$, 15\% of which
lies in the interarm regions.

\item{} Streaming motions due to density waves are evident along the minor and major
axis. The magnitude of the streaming motions is
generally very high, 60-150 \kms. Streaming, however, is absent on 
the NE arm beginning at the eastern minor axis until the arm crosses the major axis. 
Our maps offer, for the first time, sufficient resolution to detect structure in the molecular
streaming motions. Both the radial and tangential velocity components show
steep gradients in qualitative accordance with predictions by the density wave models of Roberts
and Stewart (1987). The region IC1 (in figure 6a), for example, clearly delineates the shock structure
formed on the southern
arm crossing of the major axis.{\it Our data thus support the presence of shocks in the arms of M51}.

\item{} Streaming motions (at least the tangential ones) vary in structure along the
spiral arms. The velocity gradients become generally steeper {\it between} GMAs than within them;
gradients in places reach 1000 \kms\ kpc$^{-1}$ over a small region of 100-200 pc.
Velocity gradients in GMAs are quite variable from only 60 \kms\ kpc$^{-1}$ in one GMA to several hundred
\kms\ kpc$^{-1}$ in another. 
Therefore, {\it the notion that shear
is generally reduced by streaming motions in spiral arms will have to be modified. Of the three GMAs
studied on the SW arm, only one enjoys a shear-reduced environment.} 
Conditions close to solid body rotation do occur, but apparently not everywhere in the arms.
Future statistical studies of cloud stability on the arms will have to take the varying
degree of shear into account instead of a general assumption that shear is generally low.

\item{} There is evidence of bifurcations in the 
molecular spiral arms. Emerging from these bifurcations are two chains
of molecular clouds, B1 and B2. These features may be
evidence of a secondary compression in the density wave caused
by the ultraharmonic 4/1 resonance.
We suggest that the arm bifurcations occur near the 4/1 resonance of the spiral pattern,
which is then placed between 3.2 and 3.8 kpc from the galactic center. 
This location of the resonance is consistent with results by Rand (1993b). 
The 4/1 resonance is well inside the M51 spiral pattern and this
supports the suggestion that M51 has at least two spiral systems.

\item{} Other interarm features found are a short chain of low-dispersion clouds midway between
the main and secondary arms and several molecular spurs protruding from
the main spiral arms, in particular on the western side of the map. Tangential interarm streaming motions
are detected along two of the most prominent spurs. The
agreement with the RS prediction is good; the tangential velocity decreases just before the gas enters the
arm and then increases steeply in the arm. The gradient of the interarm tangential streaming 
is $\approx 100$ \kms kpc$^{-1}$.

\item{} There is evidence in our highest resolution map (with robust weighting) for a
central molecular spiral inside the claimed location of the ILR; it is consistent with the K-band
spiral found by Zaritsky, Rix and Rieke (1993).

\item{} We find evidence of non-circular motions
in the inner 20$''$ of M51. The magnitude of these deviations is 20-30 \kms\ and they are consistent
with gas on elliptical orbits in a bar. Our data, however, show no continuous gas structure that
fills the central bar.

\item{} We find an unusual structure, an expansion in the NE molecular arm at 25$''$ radius
(1.2 kpc) southeast of the center. This broadening occurs right after
the end of the NE arm at the Inner Lindblad Resonance (ILR). 
Multiple-peak spectra, velocity twists and structures with apparently
high velocity dispersion are associated with this feature. We speculate that the disturbances in
feature C are the result of high velocity collisions between
molecular clouds in different orbits.

\end{itemize}

\acknowledgements
\sskip
We thank Magnus Thomasson for useful discussions,
Gustaf Rydbeck for estimating integrated CO fluxes from his single dish map, and
Martin Shepherd for help and advice with the AIPS and DIFMAP software packages.
We are also grateful for helpful comments and suggestions by the referee,
R. Rand. We furthermore thank S. Vogel for useful discussions of the B1 features.
The OVRO mm-array is supported in part by NSF grant AST 9314079, and 
by the K.T. and E.L. Norris Foundation.

\clearpage

\figcaption{{it a)} CO 1--0 integrated intensity map, naturally weighted with synthesized beam 
$3.''95 \times 3.''27$ and BPA=-42.8$^{\circ}$.
Contour levels are 1.1 Jy beam$^{-1}$ \kms\ $\times$ (1.0, 2.5, 4.0, ..., 17.5, 19.0). The peak flux
is 29.04 Jy beam$^{-1}$ \kms\
at $\alpha= 13^{h} 27^{m} 43^{s}.715$ ; $\delta= +47^{\circ} 26' 52''.25$ in the M1 arm. The lowest contour 
is at the 3$\sigma$ level. The total integrated flux in the map is $2.7 \times 10^3$ Jy \kms.
The solid lines mark the secondary arms B1 and B2 and the dashed lines mark the string of
interarm D  clouds (see text). Arrows indicate other features discussed in the text.
The outer map cutoff is indicated with a dashed line. This cutoff is approximately 5$''$ inside the
maps outer primary beam half power points.
{\it b)} CO 1--0 integrated intensity map robustly weighted with synthesized beam $2.''88 \times 2.''11$ 
and BPA=-80.7$^{\circ}$.
Contour levels are 0.4 Jy beam$^{-1}$ \kms\ $\times$ (1.0, 4.0, 8.0, ..., 32). The peak flux is 18.64
Jy beam$^{-1}$ \kms\ at the same position as for the naturally weighted map. The lowest contour 
is at the 3$\sigma$ level. The total integrated flux in the map is $2.0 \times 10^3$ Jy \kms.
The outer map cutoff is indicated with a dashed line.
{\it c)} Position-Velocity cut through the M1 and B1 features, at PA 242$^{\circ}$. Zero velocity 
is at 472 \kms. }

\figcaption{{\it a)} The CO radial brightness distribution for the naturally weighted map which shows
the brightness for each radial bin out to a radius of 95$''$ (corrected for inclination).  
For the conversion factor
used in this paper, Jy beam$^{-1}$ \kms\ translates to gas surface density.  
{\it b)} Intensity cut along the M1 and M2 arms. The resolution
has been smoothed to 10$''$.}

\figcaption{The CO 1--0 velocity field on the naturally weighted map. The scale range
from 350 \kms\ (blue) to 550 (red) \kms, the contour spacing is 12.5 \kms.}

\figcaption{The CO 1--0 one-dimensional dispersion map. The scale ranges
from 4 (blue) to 22 (red) \kms. }

\figcaption{The cloud density (upper panel) and velocity componets (tangential (middle
panel) and radial (lower panel)) plotted vs. spiral phase, from RS dissipative simulations, case M.
For a tightly wound spiral, a cut through spiral phase is roughly equivalent to a cut
in radius. This particular figure is at a time step 1170 Myr.}

\figcaption{ {\it a)} Velocity field ($2.''5$ resolution) overlayed on the integrated intensity
map (grayscale) in the region where arm M1 crosses the major axis. Cloud complexes A8,A9,A10
are indicated in the figure. The contour spacing is 5 \kms, the grayscale ranges from 0 to 12 Jy beam$^{-1}$
\kms. {\it b)} Velocity field ($2.''5$ resolution)
overlayed on the integrated intensity
map (grayscale) in the region where arm M2 crosses the major axis. Cloud complexes A1, A2
are indicated in the figure. The contour spacing is 5 \kms, the grayscale ranges from 0 to 8 Jy beam$^{-1}$
\kms. {\it c)} Position-Velocity cut through
the A1 GMA and the S1 spur feature, at PA 90$^{\circ}$. The velocity decreases (on the blueshifted side
of the galaxy this translates to a redshift) right before the gas enters the
arm and then increases (appears as a decrease on the blueshifted side) steeply in the arm. The expected
(Roberts 1993) peak in dispersion right before the arm is also present
{\it d)} Position-Velocity cut along the minor axis. {\it e)} Blow-up of the previous figure showing
only the cut through the western M1 arm.
Indicated as a dashed line is the Roberts and Stewart (1987) model of radial streaming motion.}

\figcaption{Blow-up of the central region of the robustly weighted moment map, figure 1b. 
K-band spiral arms are indicated with dashed lines.}

\figcaption{Inner region of the residual velocity map after a rotation curve has been
removed. Solid lines indicate redshifted velocities with respect to the model
(${v \over v_{\rm max}}=1 - e^{-ln(100 {R \over R_{\rm max}})}$). Contour spacing is 10 \kms.
Grayscale goes from 10 to 40 \kms\ residual velocities. Regions of blueshifted deviations
from the model are indicated with patterns: vertical lines indicate velocities
$\lapprox -30$ \kms; checks indicate -30 --- -20 \kms; horizontal lines indicate -20 --- -10 \kms.
White indicates regions of no deviation from circular velocities.
The greatest deviations from circular rotation
are found in two $10'' \times 20''$ regions west (redshifted) and east (blueshifted) of the center; 
the small spurs (or, rather fractions of spurs) protruding from the western arm also shows deviations
from circular velocities by up to 40 \kms; there is also a redshifted region just north of the blueshifted
region to the east. The nuclear disk is redshifted with respect to the systemic velocity.}

\figcaption{a) Sample spectrum from the southern edge of the C feature at position
$\alpha$: 13$^{\rm h}27^{\rm m}47^{\rm s}.608$, $\delta$: $47^{\circ}26'50''25$.
b) Sample spectrum from the northern edge of the C feature at position
$\alpha$: 13$^{\rm h}27^{\rm m}47^{\rm s}.806$, $\delta$: $47^{\circ}27'04''25$.}

\def\fs{\hbox{$.\!\!^{\rm s}$}}
\def\farcs{\hbox{$.\!\!^{\prime\prime}$}}

\begin{deluxetable}{lc}
\tablecaption{M51:\ \ Adopted Properties}
\tablecolumns{2}
\tablewidth{6.2in}
\tablehead{
\colhead{Parameter}&
        \colhead{Value}}
\startdata
Center position (1950.0)\tablenotemark{a} & $\alpha$: 13$^{\rm h}27^{\rm m}46^{\rm s}.327$\nl
\nodata & $\delta$: $47^{\circ}27'10''25$\nl
Morphological type\tablenotemark{b} & Sbc\nl
Systemic Velocity (hel)\tablenotemark{c} & 464 \kms\nl
Distance\tablenotemark{d} & 9.6 Mpc\nl
Position angle\tablenotemark{c} & 170$^{\circ}$\nl
Inclination\tablenotemark{c} & 20$^{\circ}$\nl
Linear resolution on galaxy (robust) &
$134 \times 98$ pc\nl
Linear resolution on galaxy (natural) &
$184 \times 152$ pc\nl
Adopted conversion factor\tablenotemark{e} & $N$(H$_2$)/$I$(CO)=$2.3 \times 10^{20}$ $\cmmt$
(K \kms)$^{-1}$
\enddata
\tablenotetext{a}{Position of radio continuum peak (Ford \etal\ 1985)}
\tablenotetext{b}{Sandage and Tammann 1981}
\tablenotetext{c}{Tully 1974a}
\tablenotetext{d}{Sandage and Tammann 1975}
\tablenotetext{e}{A standard Galactic \twco\ luminosity to H$_2$ mass ratio (Strong \etal\ 1988).
For observed CO fluxes, this translates to
$M({\rm H}_2)=0.84 \times 10^4 S \Delta v D^2$ M$_{\odot}$ ($D$ is the distance in Mpc; $S
\Delta v$ is the integrated \twco\ 1--0 line flux in Jy \kms)}.

\end{deluxetable}

\begin{deluxetable}{cll}
\tablecaption{M51:\ \ Map Pointing Centers}
\tablecolumns{3}
\tablewidth{4.2in}
\tablehead{
\colhead{Pointing}&
	\colhead{$\alpha$(1950.0)}&
        \colhead{$\delta$(1950.0)}}
\startdata
1 &  13$^{\rm h}27^{\rm m}49^{\rm s}.285$ &  $47^{\circ}28'10''25$\nl
2 &  13$^{\rm h}27^{\rm m}46^{\rm s}.327$ &  $47^{\circ}28'10''25$\nl
3 &  13$^{\rm h}27^{\rm m}43^{\rm s}.369$ &  $47^{\circ}28'10''25$\nl
\cr
4 &  13$^{\rm h}27^{\rm m}50^{\rm s}.764$ &  $47^{\circ}27'40''25$\nl
5 &  13$^{\rm h}27^{\rm m}47^{\rm s}.806$ &  $47^{\circ}27'40''25$\nl
6 &  13$^{\rm h}27^{\rm m}44^{\rm s}.848$ &  $47^{\circ}27'40''25$\nl
7 &  13$^{\rm h}27^{\rm m}41^{\rm s}.890$ &  $47^{\circ}27'40''25$\nl
\cr
8 &  13$^{\rm h}27^{\rm m}52^{\rm s}.242$ &  $47^{\circ}27'10''25$\nl
9 &  13$^{\rm h}27^{\rm m}49^{\rm s}.285$ &  $47^{\circ}27'10''25$\nl
10 &  13$^{\rm h}27^{\rm m}46^{\rm s}.327$ &  $47^{\circ}27'10''25$\tablenotemark{a}\nl
11 &  13$^{\rm h}27^{\rm m}43^{\rm s}.369$ &  $47^{\circ}27'10''25$\nl
12 &  13$^{\rm h}27^{\rm m}40^{\rm s}.412$ &  $47^{\circ}27'10''25$\nl
\cr
13 &  13$^{\rm h}27^{\rm m}50^{\rm s}.764$ &  $47^{\circ}26'40''25$\nl
14 &  13$^{\rm h}27^{\rm m}47^{\rm s}.806$ &  $47^{\circ}26'40''25$\nl
15 &  13$^{\rm h}27^{\rm m}44^{\rm s}.848$ &  $47^{\circ}26'40''25$\nl
16 &  13$^{\rm h}27^{\rm m}41^{\rm s}.890$ &  $47^{\circ}26'40''25$\nl
\cr
17 &  13$^{\rm h}27^{\rm m}49^{\rm s}.285$ &  $47^{\circ}26'10''25$\nl
18 &  13$^{\rm h}27^{\rm m}46^{\rm s}.327$ &  $47^{\circ}26'10''25$\nl
19 &  13$^{\rm h}27^{\rm m}43^{\rm s}.369$ &  $47^{\circ}26'10''25$
\enddata
\tablenotetext{a}{Map center position}

\end{deluxetable}

\begin{deluxetable}{lccc}
\tablecaption{M51:\ \ Tangential streaming and shear}
\tablecolumns{4}
\tablewidth{5.2in}
\tablehead{
\colhead{Region}&
        \colhead{${dV_{\rm tan} \over dr}$\tablenotemark{a}}&
        \colhead{$A(R)$\tablenotemark{b}}&
        \colhead{$t\tablenotemark{c}_{\rm shear}$}\nl
\colhead{}&
	\colhead{(\kms\ kpc$^{-1}$)}&
	\colhead{(\kms\ kpc$^{-1}$)}&
	\colhead{($10^7$ yr)}}
\startdata
GMAs:\nl
A2 & 250 & -85 & 1.1\nl
A1 & 115 & -18 & 5.3\nl
A8 & 250  & -85 & 1.1\nl
A9 & 65  & 8 & 12\nl
A10 & 700 & -310 & 0.3\nl
\cr
Inter-cloud regions:\nl
IC1 & 800 - 1300 & -360 --- -610 & 0.3 --- 0.2\nl
IC2 & 600 & -260 & 0.4\nl
IC3 & 400 & -160 & 0.6

\enddata
\tablenotetext{a}{Velocity gradient in the tangential streaming motions}
\tablenotetext{b}{Oort constant at radius $R$, $A(R)={1 \over 2} ({v_c \over R} - {dv_c \over dR})$
where $v_c$ is the circular velocity. At $R$=2.7 kpc ${v_c \over R}$ is 80 \kms\ kpc$^{-1}$.}
\tablenotetext{c}{For pure differential rotation, $t_{\rm shear}$ is $2.4 \times 10^7$ yr
at $R$=2.7 kpc.}

\end{deluxetable}

\begin{deluxetable}{llll}
\tablecaption{M51:\ \ Position of resonances\tablenotemark{a}}
\tablecolumns{4}
\tablewidth{7.2in}
\tablehead{
\colhead{Reference\tablenotemark{b}}&
        \colhead{ILR}&
        \colhead{corotation}&
        \colhead{Comments}}
\startdata
Garc\'{\i}a-Burillo\tablenotemark{1} & 1.7  & 7.4 & assumes single pattern\nl
Tully\tablenotemark{2} & 1.4 & 5.7 - 7.1 & two spiral patterns\nl
Elmegreen\tablenotemark{3}  & \nodata & 6.0 & two spiral pattern\nl
Vogel\tablenotemark{4}  & \nodata & 6.0 & two spiral patterns
\enddata
\tablenotetext{a}{All data has been corrected to a distance $D$=9.6 Mpc}
\tablenotetext{b}{(1)Garc\'{\i}a-Burillo \etal\ (1993b); (2)Tully 1974b; (3)Elmegreen,
Elmegreen and Seiden (1989); (4) Vogel \etal\ (1993)}

\end{deluxetable}

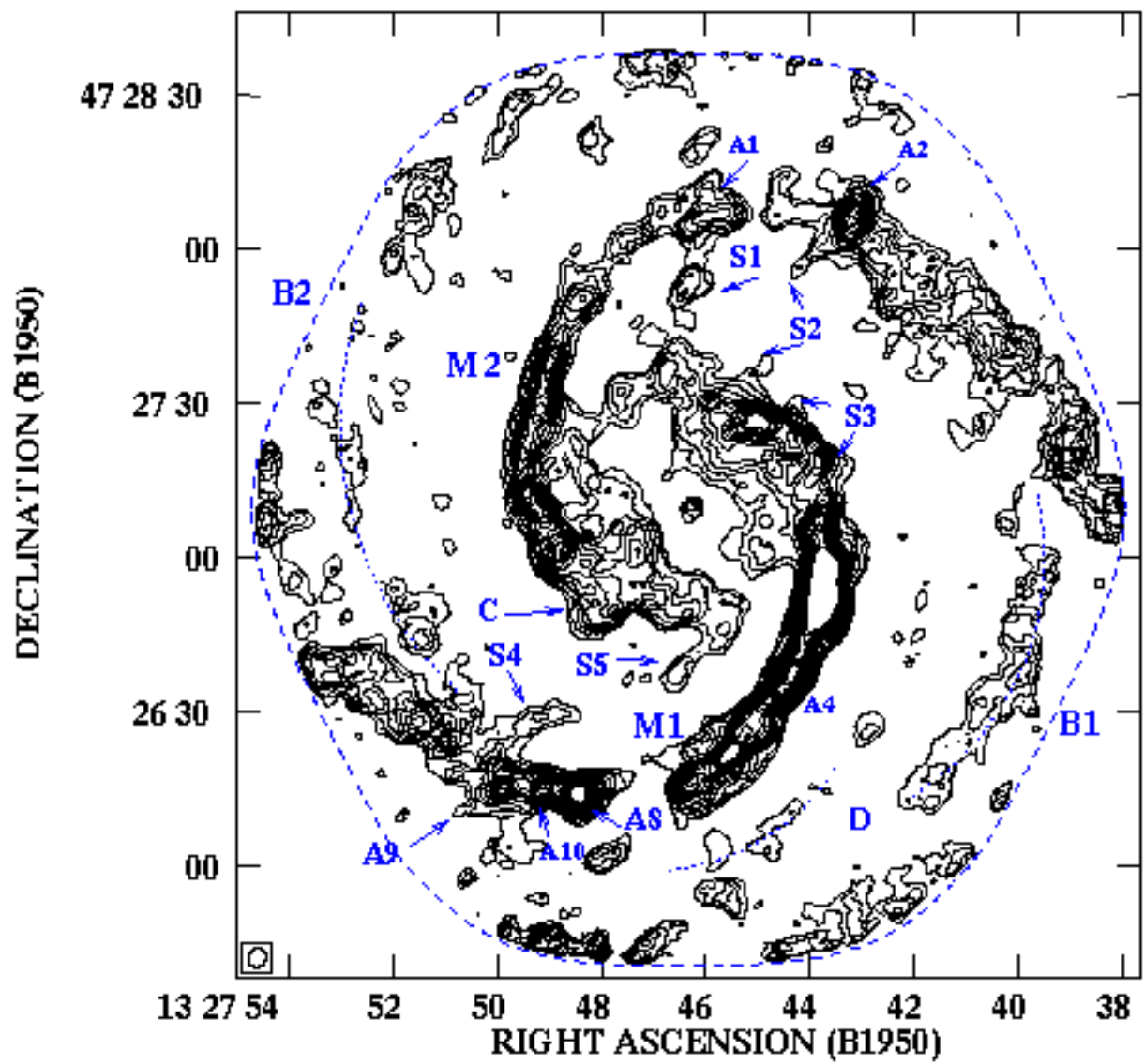
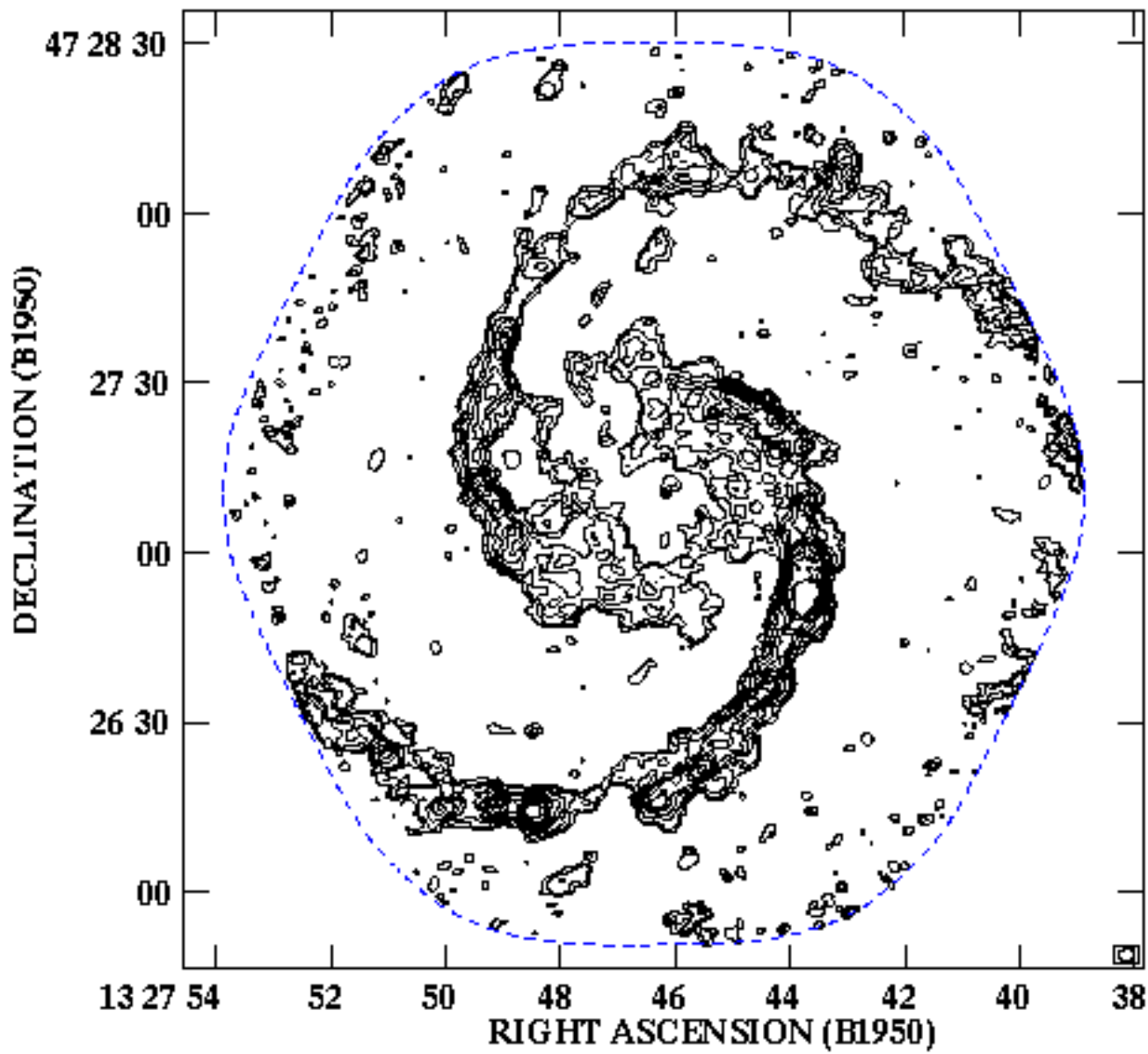
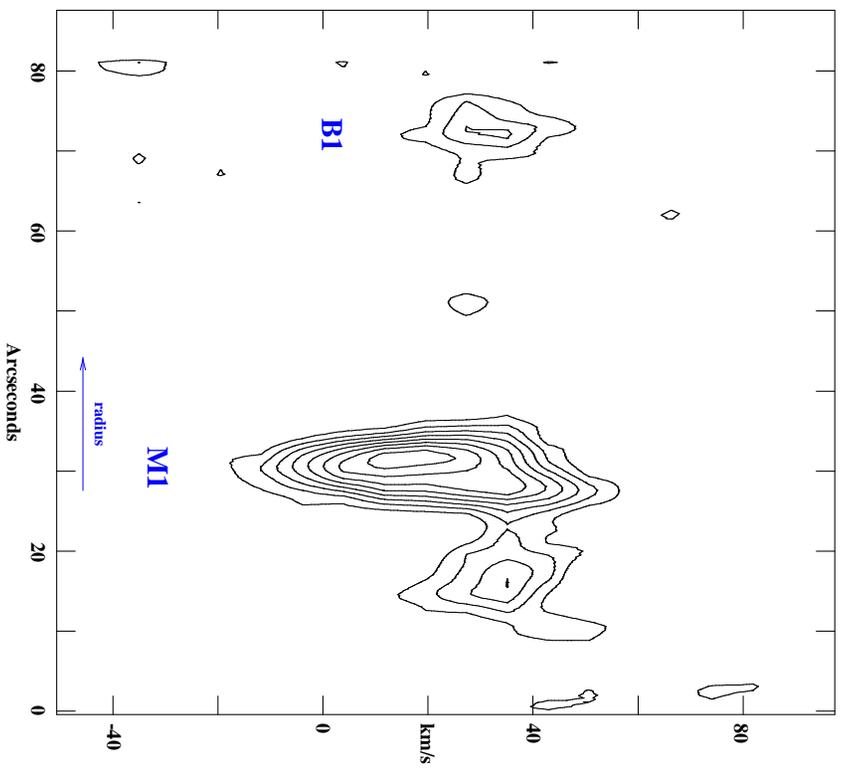
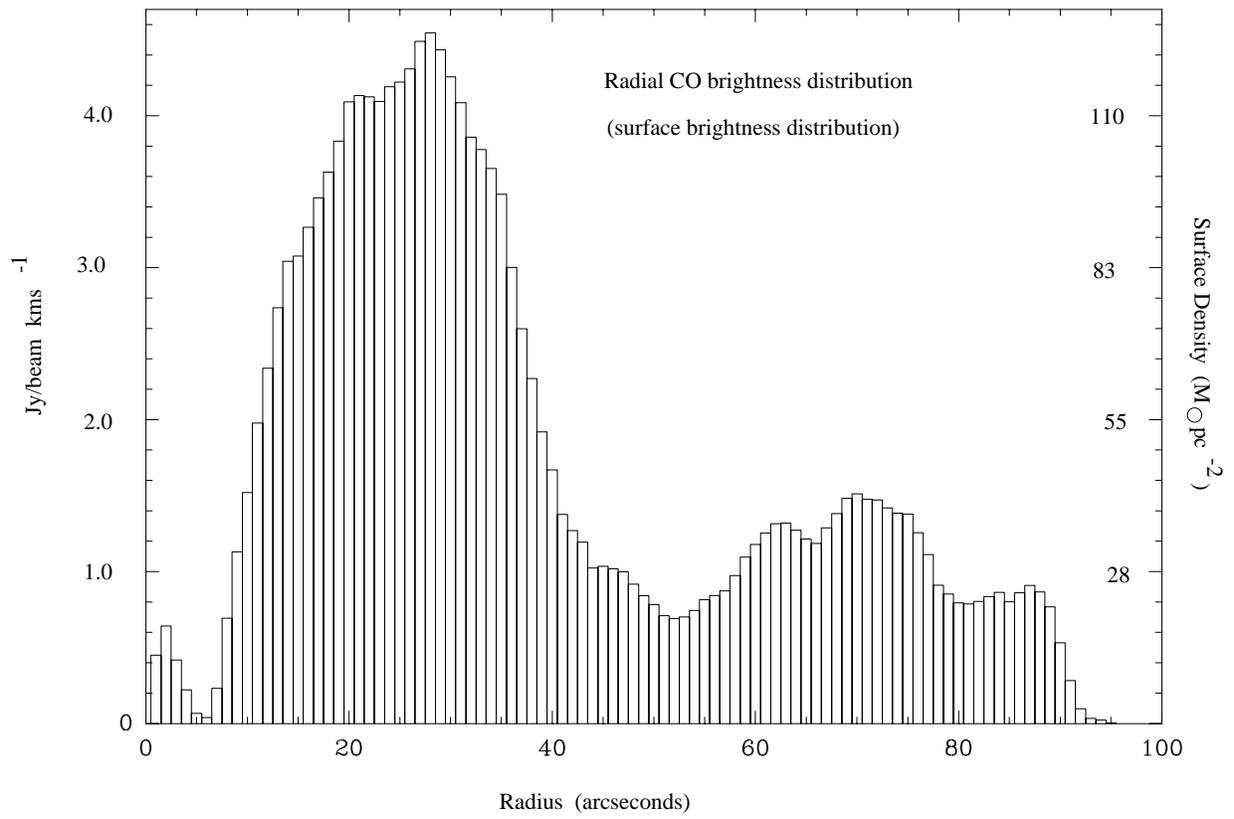
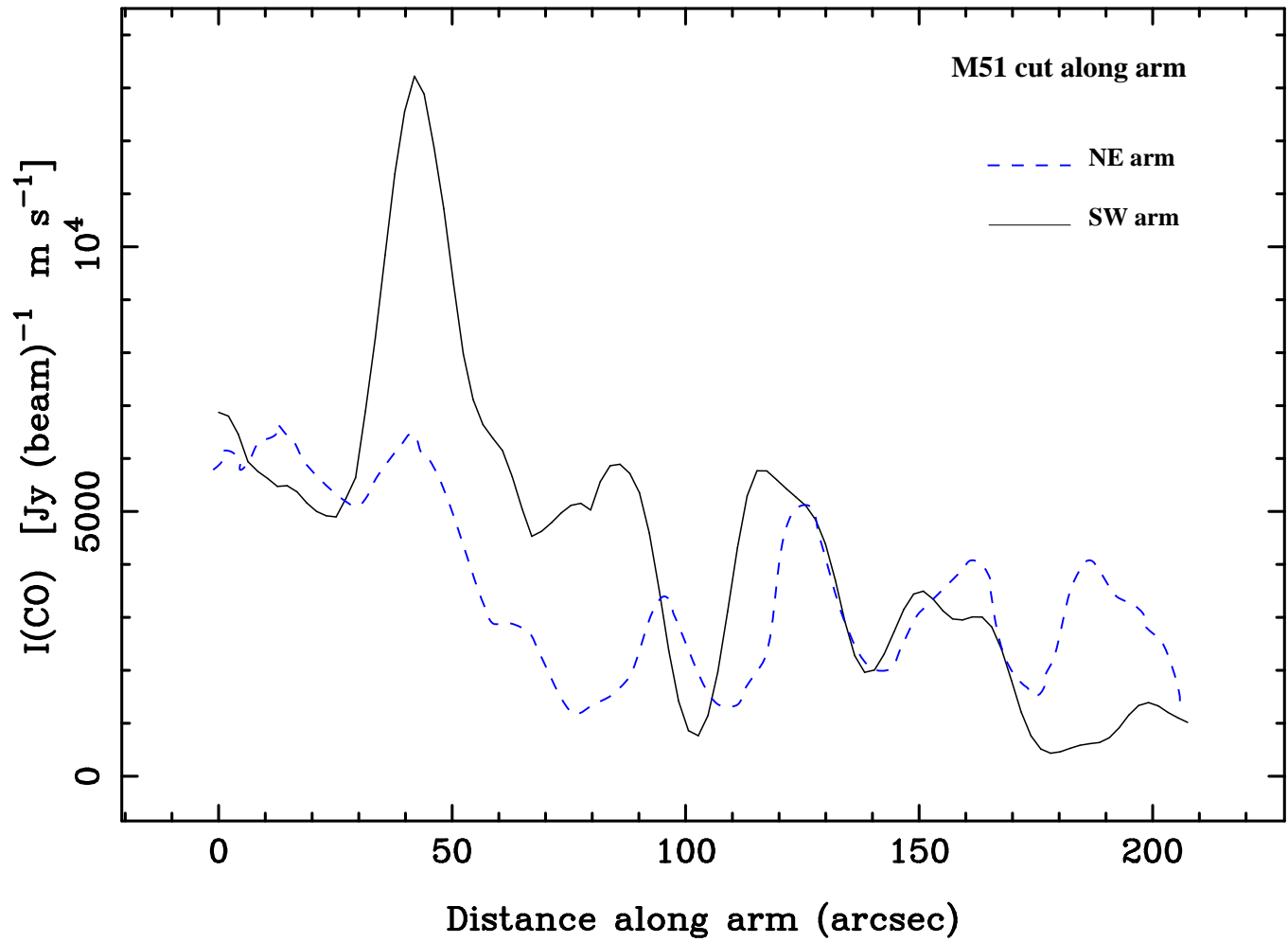
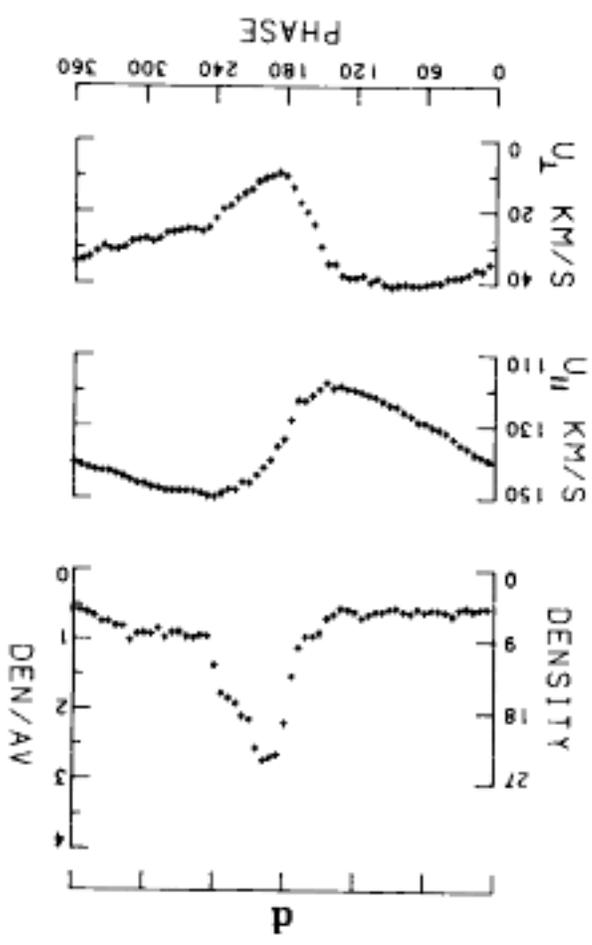
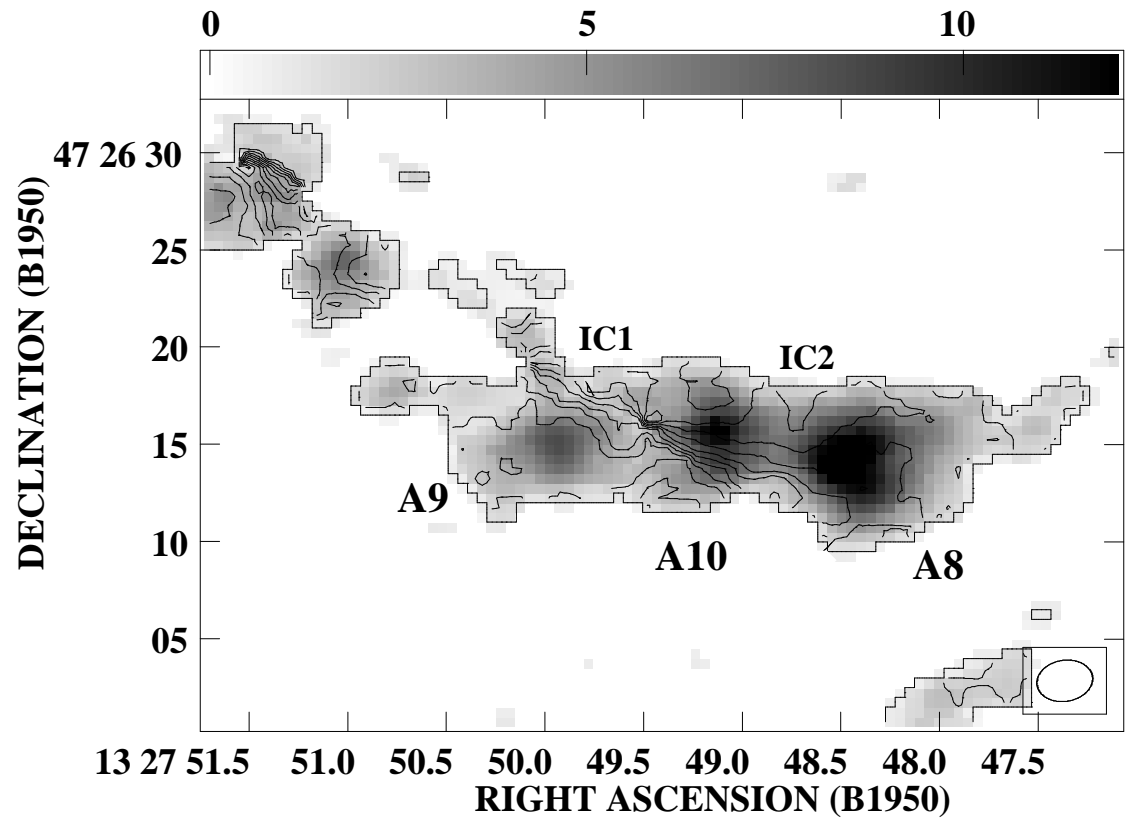
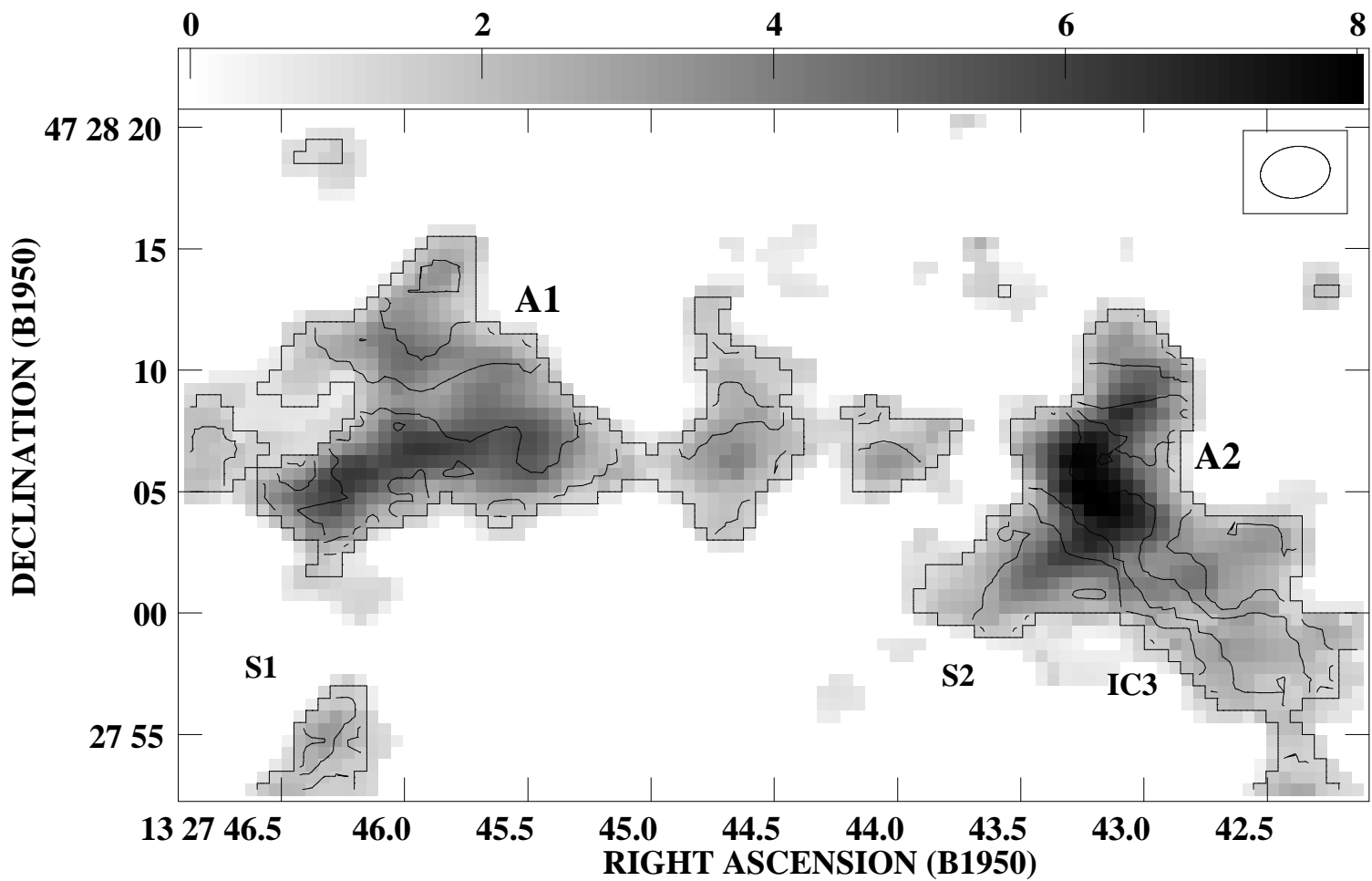
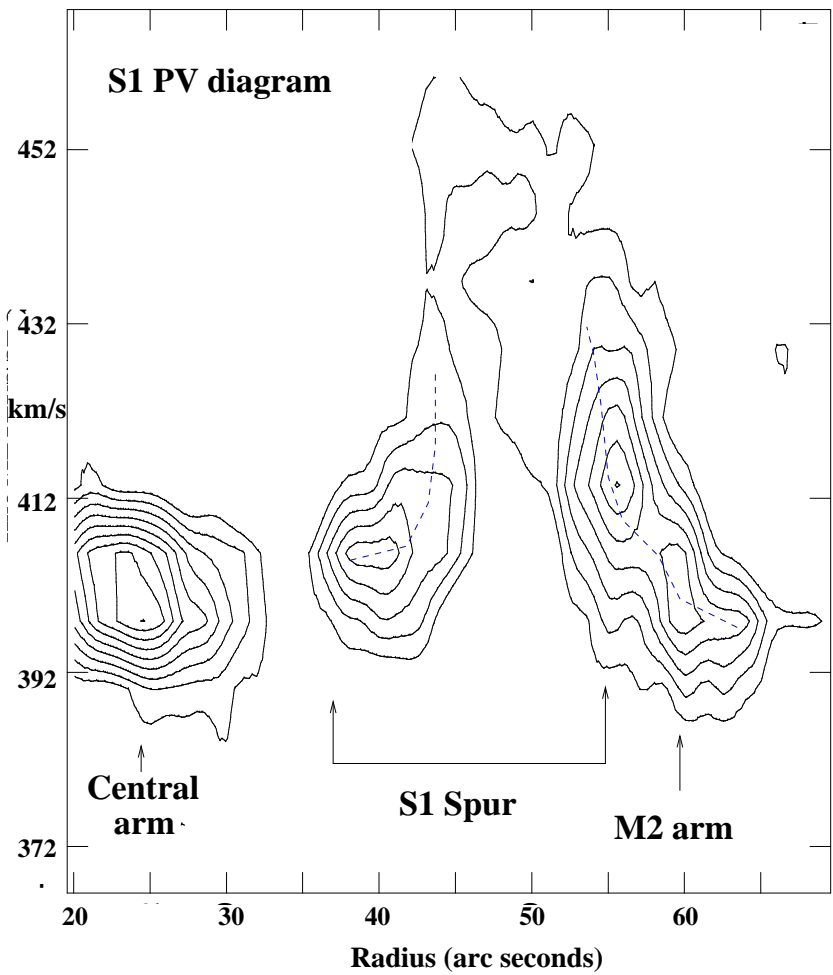
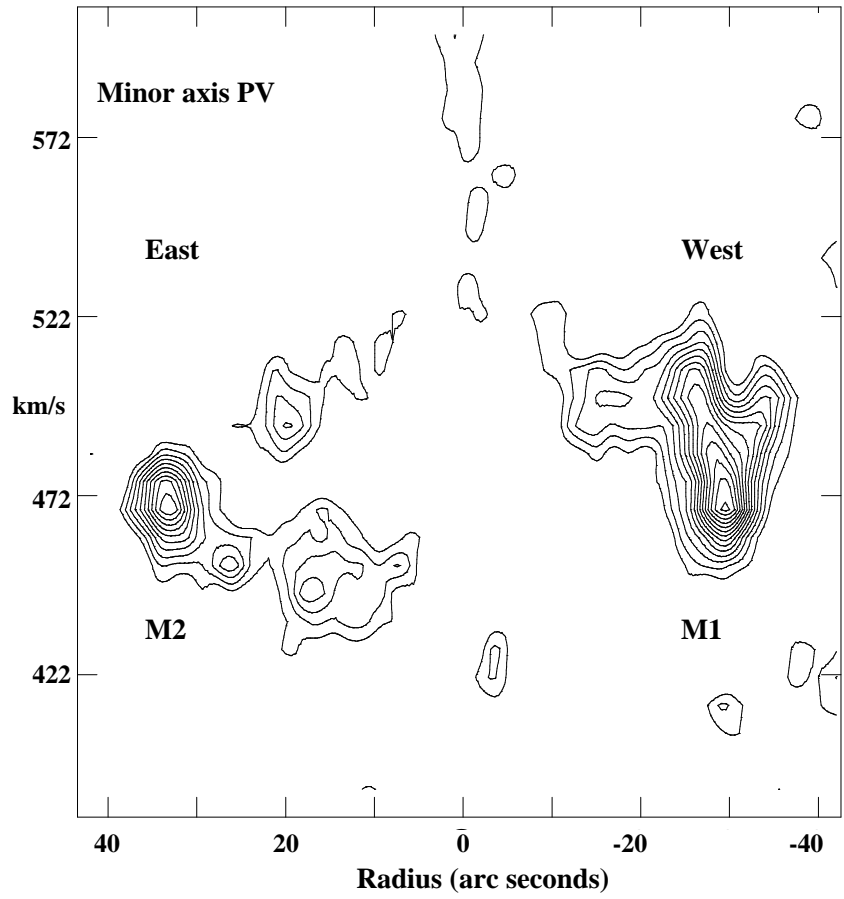
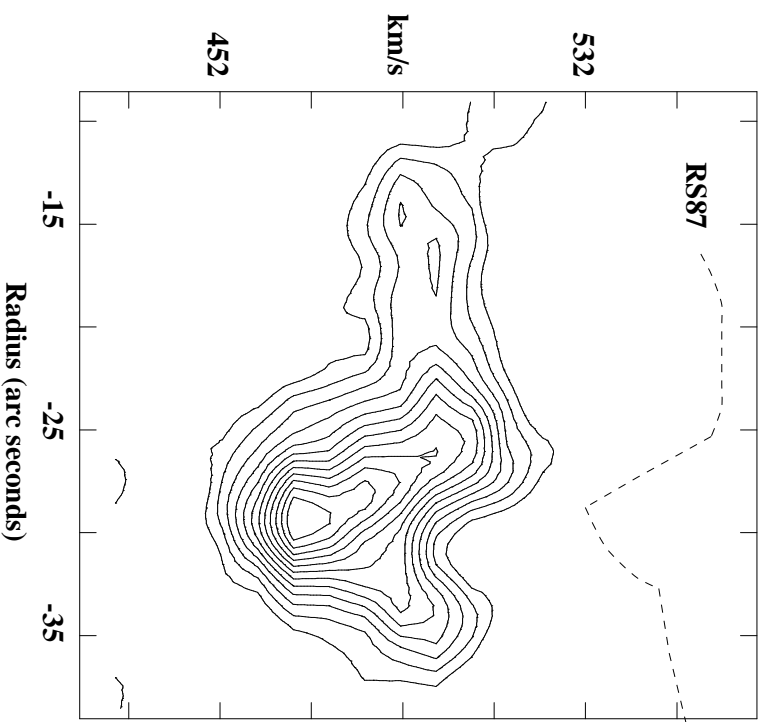
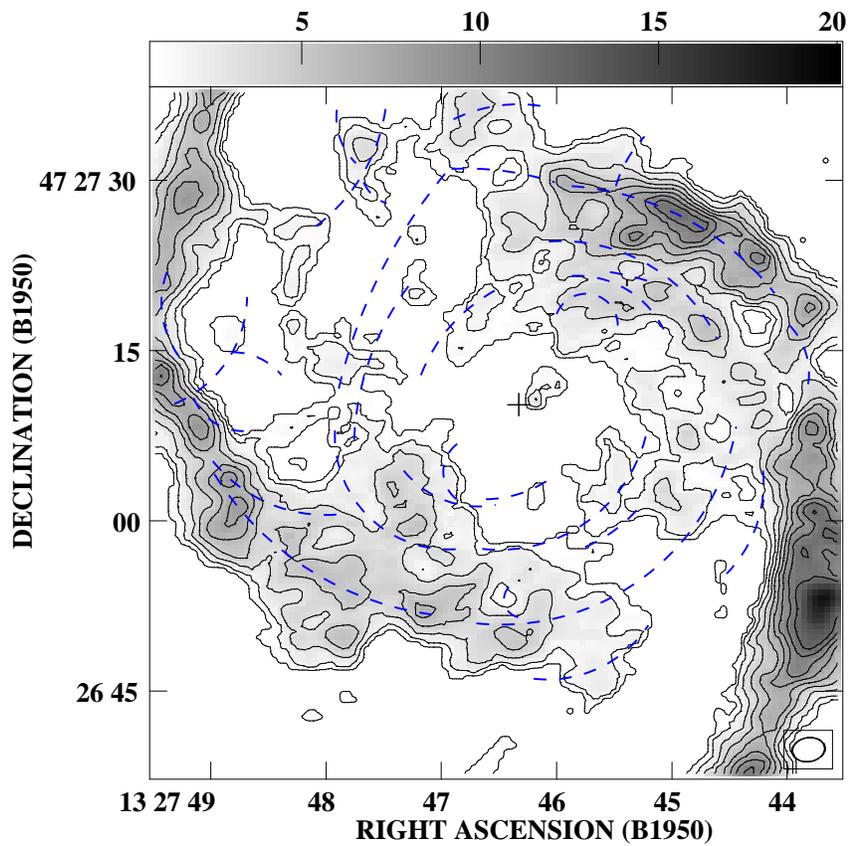
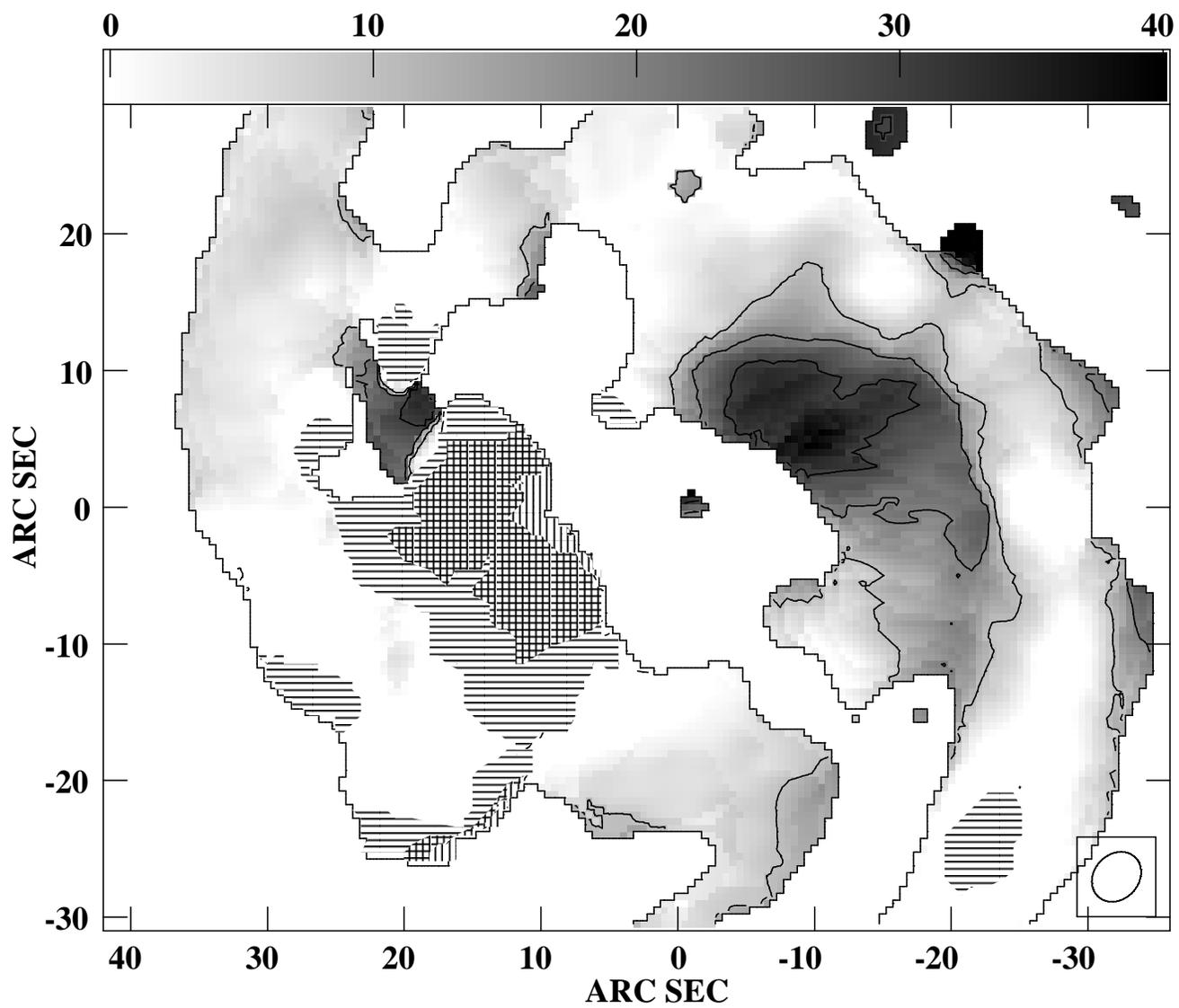
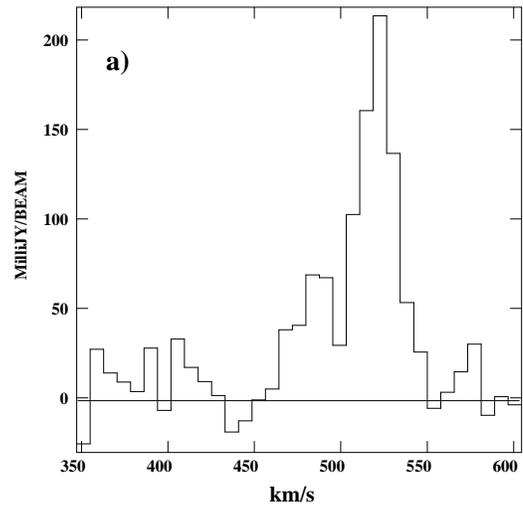
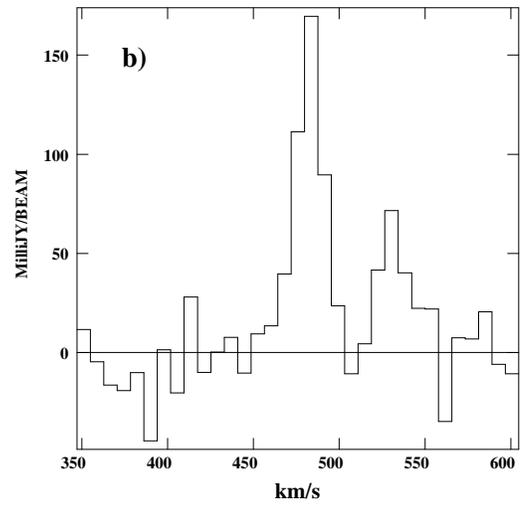
\end{document}